\documentclass{elsart}

\usepackage{graphicx}
\usepackage{ifthen}
\usepackage{xspace}
\usepackage{pst-text}
\usepackage{pst-tree}
\usepackage{pst-grad}
\usepackage{pst-node}
\usepackage{amssymb}
\usepackage{amsmath}
\usepackage{cite}                 %% Sort and compress citations 
\usepackage{subfigure}
\usepackage{url}
\newcommand{\gpop}[1]{\Toval{\textcolor{red}{#1}}}
\newcommand{\gpbreakop}[1]{\Toval[linestyle=dashed]{\textcolor{red}{#1}}}
\newcommand{\gpterm}[1]{\Toval{\textcolor{blue}{#1}}}

\providecommand{\eg}{\emph{e.g.}\xspace}
\providecommand{\Eg}{\emph{E.g.}\xspace}
\providecommand{\ie}{\emph{i.e.}\xspace}

\newcommand{\cer}{\v{C}erenkov\xspace}

\newcommand{\mc}{Monte Carlo\xspace}

\newcommand{\dcs}{doubly Cabibbo suppressed\xspace}

\newcommand{\cf}{Cabibbo favored\xspace}
\newcommand{\gp}{genetic programming\xspace}
\newcommand{\GP}{Genetic Programming\xspace}
\newcommand{\Gp}{Genetic programming\xspace}
\newcommand{\first}{1$^\text{st}$\xspace}
\newcommand{\second}{2$^\text{nd}$\xspace}

\newcommand{\lsig}{\ensuremath{\ell/\sigma_\ell}\xspace}

\newcommand{\chisq}{\ensuremath{\chi^2}\xspace}

\newcommand{\dwpik}{\ensuremath{\Delta W(\pi \kaon)}\xspace}

\newcommand{\picon}{\ensuremath{\pi_{\mathrm{con}}}\xspace}

\newcommand{\mathsl}[1]{\mbox{\textsl{#1}}}

\newcommand{\meson}[1]{\ensuremath{\mathsl{#1}}}

\newcommand{\muon}{\ensuremath{\mu}\xspace}

\newcommand{\pion}{\ensuremath{\pi}\xspace}
\newcommand{\piplus}{\ensuremath{\pion^+}\xspace}
\newcommand{\piminus}{\ensuremath{\pion^-}\xspace}

\newcommand{\kaon}{\ensuremath{\meson{K}}\xspace}

\newcommand{\kplus}{\ensuremath{\kaon^+}\xspace}

\newcommand{\dmeson}{\ensuremath{\meson{D}}\xspace}
\newcommand{\dplus}{\ensuremath{\dmeson^{+}}\xspace}

\newcommand{\dsplus}{\ensuremath{\dmeson_{s}^{+}}\xspace}

\newcommand{\bmeson}{\ensuremath{\meson{B}}\xspace}

\newcommand{\kpipi}{\ensuremath{\dplus \to  \kaon^- \piplus \piplus}\xspace}
\newcommand{\kpipidcsd}{\ensuremath{\dplus \to  \kaon^+ \piplus \piminus}\xspace}

\newcommand{\mevcc}{\ensuremath{\mathrm{MeV}/c^2}\xspace}

\newcommand{\gevcc}{\ensuremath{\mathrm{GeV}/c^2}\xspace}
\newcommand{\gevc}{\ensuremath{\mathrm{GeV}/c}\xspace}
\newcommand{\gev}{\ensuremath{\mathrm{GeV}}\xspace}

\newcommand{\secref}[1]{Section~\ref{sec:#1}}

\newcommand{\eqnref}[1]{Eq.~(\ref{eqn:#1})}
\newcommand{\figref}[1]{Figure~\ref{fig:#1}}

\newcommand{\tabref}[1]{Table~\ref{tab:#1}}

\newcommand{\figlabel}[1]{\label{fig:#1}}
\newcommand{\eqnlabel}[1]{\label{eqn:#1}}
\newcommand{\tablabel}[1]{\label{tab:#1}}
\newcommand{\seclabel}[1]{\label{sec:#1}}

\newcommand{\routine}[1]{\textsf{\textbf{#1}}}

\renewcommand{\gpop}[1]{\Toval{#1}}
\renewcommand{\gpterm}[1]{\Toval{#1}}
\renewcommand{\gpbreakop}[1]{\Toval[linestyle=dashed]{#1}}
\renewcommand{\figref}[1]{Fig.~\ref{fig:#1}}

\begin{document}                                                 
%\begin{fmffile}{gp_memo_feyn}

\begin{frontmatter}

\title{Application of Genetic Programming to High Energy Physics Event Selection}

%\author{THE FOCUS COLLABORATION}
The FOCUS Collaboration

\author[ucd]{J.~M.~Link}
\author[ucd]{P.~M.~Yager}
\author[cbpf]{J.~C.~Anjos}
\author[cbpf]{I.~Bediaga}
\author[cbpf]{C.~Castromonte}
\author[cbpf]{C.~G\"obel}
\author[cbpf]{A.~A.~Machado}
\author[cbpf]{J.~Magnin}
\author[cbpf]{A.~Massafferri}
\author[cbpf]{J.~M.~de~Miranda}
% ...

\address[ucd]{University of California, Davis, CA 95616}
\address[cbpf]{Centro Brasileiro de Pesquisas F\'isicas, Rio de Janeiro, RJ, Brasil}
\address[cinv]{CINVESTAV, 07000 M\'exico City, DF, Mexico}
\address[cu]{University of Colorado, Boulder, CO 80309}
\address[fnal]{Fermi National Accelerator Laboratory, Batavia, IL 60510}
\address[fras]{Laboratori Nazionali di Frascati dell'INFN, Frascati, Italy I-00044}
\address[ugj]{University of Guanajuato, 37150 Leon, Guanajuato, Mexico} 
\address[ui]{University of Illinois, Urbana-Champaign, IL 61801}
\address[iu]{Indiana University, Bloomington, IN 47405}
\address[korea]{Korea University, Seoul, Korea 136-701}
\address[kp]{Kyungpook National University, Taegu, Korea 702-701}
\address[milan]{INFN and University of Milano, Milano, Italy}
\address[nc]{University of North Carolina, Asheville, NC 28804}
\address[pavia]{Dipartimento di Fisica Nucleare e Teorica and INFN, Pavia, Italy}
\address[pr]{University of Puerto Rico, Mayaguez, PR 00681}
\address[sc]{University of South Carolina, Columbia, SC 29208}
\address[ut]{University of Tennessee, Knoxville, TN 37996}
\address[vu]{Vanderbilt University, Nashville, TN 37235}
\address[wisc]{University of Wisconsin, Madison, WI 53706}

\centerline{\small See \textrm{http://www-focus.fnal.gov/authors.html} for additional author information.}

\begin{abstract}  
We review genetic programming principles, their application to FOCUS data
samples, and use the method to study the \dcs decay \kpipidcsd  relative to its
\cf counterpart, \kpipi. We find that this technique is able to improve upon
more traditional analysis methods. To our knowledge, this is the first
application of the \gp technique to High Energy Physics data.
\end{abstract}

\end{frontmatter}

\section{Introduction}\seclabel{intro}

Genetic programming is one of a number of machine learning techniques in which
a computer program is given the elements of possible solutions to the problem.
This program, through a feedback mechanism, attempts to discover the best
solution to the problem at hand, based on the programmers definition of
success. Genetic programming differs from machine learning solutions such as
genetic algorithms and neural networks in that the form or scope of the solution is
not specified in advance, but is determined by the complexity of the problem.
We have applied this technique to the study of a \dcs branching ratio  of
\dplus since this measurement is presumed to be reasonably free of \mc related
systematic errors. To our knowledge, this is the first application of the \gp
technique to high energy physics data, although we have recently become aware
of a \mc study~\cite{Cranmer:2004gp} for the ATLAS experiment.  

First, we review the basics of \gp and explain how \gp is used on FOCUS data.   Next, we
show the results of applying \gp techniques to the \dcs decay\footnote{Throughout this
paper, charge conjugate states are implied.}  \kpipidcsd,  measure its branching ratio
relative to the \cf decay, and compare our results with a published analysis using more
conventional methods. Finally, we describe studies of systematic errors related to the
\gp method. 

\section{Introduction to Genetic Programming}\seclabel{gpintro}

We have adopted the tree representation and nomenclature of
Koza~\cite{Koza:gp1992} as our \gp model. Throughout this paper, we will use
the terms ``tree,'' ``program,'' and ``individual'' interchangeably. Other, more
general, representations also exist~\cite{Banzhaf:gp1998}. The specific
implementation used is lil-gp~\cite{www_lilgp} from the Michigan State
University GARAGe group~\cite{www_garage}. In this representation, the \GP
Framework (GPF) creates a program which consists of a series of linked nodes.
Each node takes a number of arguments and supplies a single return value. There
are two general types of nodes: functions (or operators) and terminals
(variables and constants). Functions take one or more input variables;
terminals take no inputs and supply a single value to the program. In practice,
the argument and return ``values'' can be any data structure, but in our case
they are single floating point numbers.

This series of linked nodes can be represented as a tree where the leaves
of the tree represent terminals and operators reside at the forks of the tree.
For example, \figref{tree-example} shows the representation of the function ($p
\times p) - (p + 40)$. To ``read'' trees in this fashion, one resolves the
sub-trees in a bottom-up fashion.

\begin{figure}
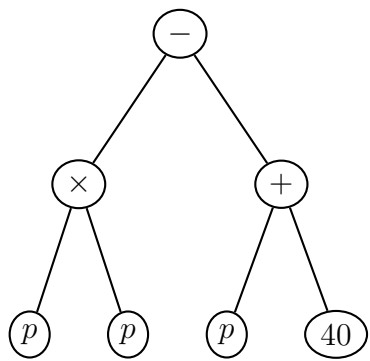

\begin{center}
\pstree{\gpop{$-$}}{
 \pstree{\gpop{$\times$}}{\gpterm{$p$}\gpterm{$p$}}
 \pstree{\gpop{$+$}}{\gpterm{$p$}\gpterm{$40$}}
}
\end{center}
\caption{Tree representation of the equation 
$(p \times p) - (p + 40)$.}
\figlabel{tree-example} 
\end{figure}

The \gp model seeks to mimic the biological processes of evolution, treating
each of these trees or programs as an ``organism.'' Through natural selection and
reproduction over a number of generations, the fitness (\ie, how well the
program solves the specified problem) of a population of organisms is improved.

The process of determining the best (or nearly best) solution to a problem in
genetic programming involves a series of steps:

\begin{enumerate}
  \item{Generate an initial population of ``programs'' to be evaluated.}
  \item{Determine the fitness of each of these programs in solving the
        problem at hand.}
  \item{Select which of these programs are allowed to contribute to the next
        generation.}
  \item{Using these selected programs, generate the next generation using biological
        models.}
  \item{Repeat steps 2--4 a number of times.}
  \item{Terminate and evaluate the best solution(s).}
\end{enumerate}

\subsection{Initial tree generation}\seclabel{tree_gen}

The initial trees to be evaluated are created in one of two ways. The
first, termed ``grow,'' randomly selects either a terminal or a function as
the first (top) node in the tree. If a function is selected, the ``child'' nodes
required by that function are again randomly selected from either functions or
terminals. Growth is stopped either when all available nodes have been filled
with terminals or when the maximum depth (measured from top to bottom) for the tree is reached (at which
point only terminals are chosen).

The second method, termed ``full,'' proceeds similarly, except only functions
are chosen at levels less than the desired depth of the initial tree and only
terminals are selected at the desired depth. Typically one specifies a range
of depths, which results in subpopulations of differing complexity.

In the initial generation, every tree in the population is guaranteed to be
unique.\footnote{This is not quite accurate. Every tree within a sub-process
is unique; we run with up to 40 sub-processes, as explained later.} Later
generations are allowed to have duplicate individuals.

For a wide range of problems it has been observed that generating half the
initial population with the grow method and the other half with various depths
of full trees provides a good balance in generating the initial population.

\subsection{Fitness evaluation}

Central to  \gp is the idea of fitness. Fitness is the measure of how well the
program, or tree, solves the problem presented by the user. The GPF we use requires
the calculation of \emph{standardized fitness} such that the best possible solution
will have a fitness of 0 and all other solutions will have positive fitness
values. The exact definition of the standardized fitness is left to the
programmer.

Another useful quantity is the adjusted fitness,
\begin{equation}
f_a(i) = \frac{1}{1+f_s(i)}
\end{equation}
where $f_a(i)$ is the adjusted fitness of the $i^\text{th}$ individual and $f_s(i)$
is the standardized fitness of the same individual. One can readily see that as
$f_s$ decreases, $f_a$ increases to a maximum of 1.0.  

\subsection{Survival selection}

To mimic the evolutionary process of natural selection, the probability that a 
particular tree will pass some of its ``genetic material,'' or instructions, on
to the next generation must be proportional in some way to the fitness of that
individual. 

In \gp, several kinds of selection are employed. The first is
``fitness-proportionate,'' in which the probability, $p_i$, of selection of the
$i^\text{th}$ individual is 
\begin{equation}
p_i = \frac{f_a(i)}{\sum_j f_a(j)}
\end{equation} 
where $j$ sums over all the individuals in a population. In this way, very fit
individuals are selected more often than relatively unfit individuals. This is
the most common method of selection in \gp.

For complicated problems which require larger population sizes,
``fitness-over-selection'' is often used. In this method, the population is
divided into two groups, a small group of ``good'' individuals and a larger
group of the remaining individuals. The majority of the time, an individual is
selected from the smaller group, using the rules for fitness-proportionate
selection. The rest of the time, the individual is selected from the larger
group, again with the same rules.

In our implementation, the small group contains the best individuals which
account for (320/population size) of the total adjusted fitness. \Eg, for a
population size of 4000, the small group contains the individuals summing to
8\% of the total adjusted fitness. Individuals are selected from this small
group 80\% of the time.

Additional types of selection are sometimes used in genetic programming. These
include ``tournament,'' in which two or more individuals are randomly chosen and
the best is selected, and ``best,'' in which the best individuals are selected in
order.

\subsection{Breeding algorithms}

The process of creating a new generation of programs from the preceding
generation is called ``breeding'' in the \gp literature. The similar sounding
term, ``reproduction,'' is used in a specific context as explained below. 

The methods used to produce new individuals are determined randomly; more
than one method can be (and is) used. Each of these methods must select one or more
individuals as ``parents.'' The parents are selected according to the methods
described above.  The GPF we are using implements three such methods: reproduction,
crossover, and mutation. Other methods, \eg permutation, also
exist~\cite{Koza:gp1992}.

\subsubsection{Reproduction}

``Reproduction'' might also be called cloning or asexual reproduction. The selected
individual is simply copied into the next generation without alteration. 

\subsubsection{Crossover}

``Crossover,'' or sexual reproduction, randomly selects two parent trees and
creates two new trees. In an attempt to mimic DNA exchange between two parents,
a node is selected on each tree, those nodes and all child nodes are removed
from the parent trees and inserted into the vacant spot in the other tree. This
process is illustrated in \figref{xover-example}.   

\begin{figure}
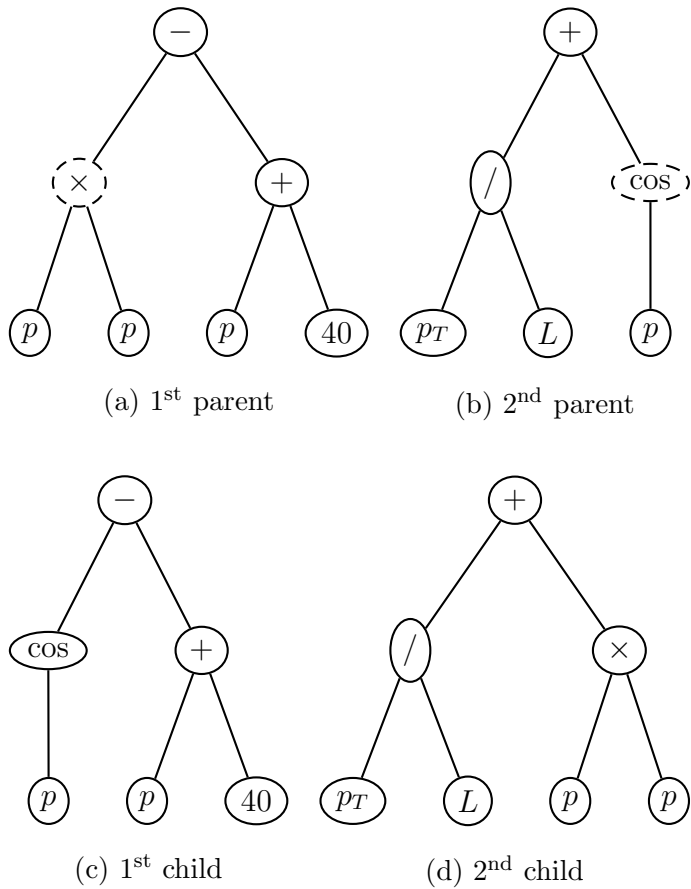

\centering
\subfigure[\first parent]{
  \pstree{\gpop{$-$}}{
   \pstree{\gpbreakop{$\times$}}{\gpterm{$p$}\gpterm{$p$}}
   \pstree{\gpop{$+$}}{\gpterm{$p$}\gpterm{$40$}}
  }
}
\subfigure[\second parent]{
  \pstree{\gpop{$+$}}{
   \pstree{\gpop{$/$}}{\gpterm{$p_T$}\gpterm{$L$}}
   \pstree{\gpbreakop{$\cos$}}{\gpterm{$p$}}
  }
}
\\
\subfigure[\first child]{
  \pstree{\gpop{$-$}}{
   \pstree{\gpop{$\cos$}}{\gpterm{$p$}}
   \pstree{\gpop{$+$}}{\gpterm{$p$}\gpterm{$40$}}
  }
}
\subfigure[\second child]{
  \pstree{\gpop{$+$}}{
   \pstree{\gpop{$/$}}{\gpterm{$p_T$}\gpterm{$L$}}
   \pstree{\gpop{$\times$}}{\gpterm{$p$}\gpterm{$p$}}
  }
}
\caption{Example of crossover. Shown are the two parents (a) and (b) and the two
resulting children (c) and (d). The crossover points chosen are designated with
the dashed ovals.}
\figlabel{xover-example}
\end{figure}

\subsubsection{Mutation}

In mutation, a single parent is selected. A node on the tree is selected, the
existing contents and any child nodes are destroyed, and a new node (terminal
or function) is inserted in its place.  New nodes are inserted following the
rules for growing the initial trees. Both the tree and mutation point are
selected randomly. An example of mutation is shown in \figref{mutate-example}.

\begin{figure}
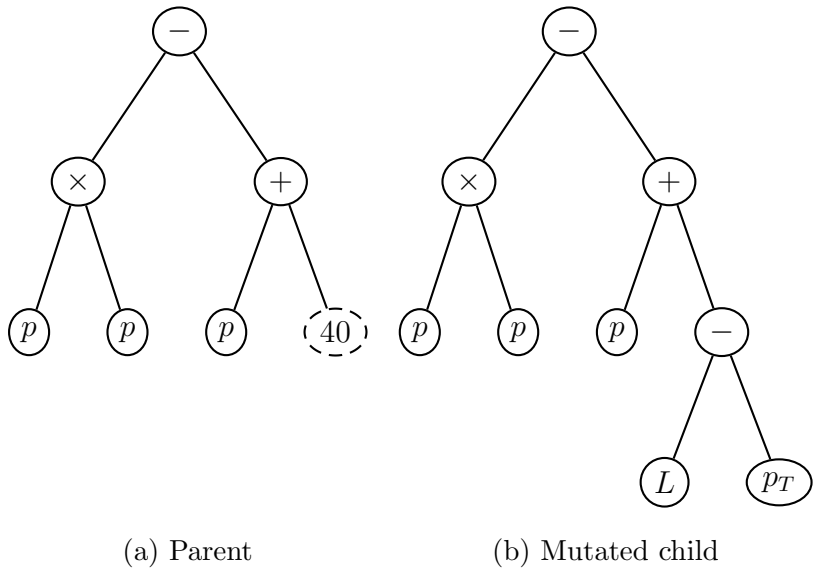

\centering
\subfigure[Parent]{\raisebox{-0.1in}[0in][2.6in]{
  \pstree{\gpop{$-$}}{
   \pstree{\gpop{$\times$}}{\gpterm{$p$}\gpterm{$p$}}
   \pstree{\gpop{$+$}}{\gpterm{$p$}\gpbreakop{$40$}}
  }
}}
\subfigure[Mutated child]{\raisebox{-0.1in}[0in][2.6in]{
  \pstree{\gpop{$-$}}{
   \pstree{\gpop{$\times$}}{\gpterm{$p$}\gpterm{$p$}}
   \pstree{\gpop{$+$}}{\gpterm{$p$}\pstree{\gpop{$-$}}{\gpterm{$L$}\gpterm{$p_T$}}}
  }
}}
\caption{Example of mutation. Shown are the parent (a) and the 
resulting child (b). The mutation point is denoted by the dashed oval.}
\figlabel{mutate-example}
\end{figure}

Typically the amount of mutation is kept small since it is exploring new
parameter space, not parameter space that has proven to be interesting already.
However, as in biological processes, mutation may be important to preserve or
increase diversity. Note that mutation beginning with the first node is equivalent to
generating a completely new individual.

\subsection{Migration}

Although not part of early \gp models, we use migration of individuals as an
important element of generating diversity. In our \emph{parallel} \gp model,
sub-populations (or ``islands'') are allowed to evolve independently  with
exchanges between sub-populations taking place every few generations. Every
$n_g$ generations, the best $n_i$ individuals from each sub-population are
copied into every other sub-population. After this copying, they may be
selected by the methods discussed above for the various breeding processes. 

This modification to the method allows for very large effective population to
be spread over a large number of computers in a network.

\subsection{Termination}
 
When the required number of generations have been run and all the individuals
evaluated, the GPF terminates. At this point, a text representation of the
best tree (not necessarily from the last generation) is written out and other
cleanup tasks are performed.

\section{Combining Genetic Programming with High Energy Physics Data}\seclabel{focusgp}

While \gp techniques can be applied to a wide variety of problems in High Energy
Physics (HEP), we illustrate the technique on the common problem of event
selection: the process of distinguishing signal events from the more copious
background processes.
Event selection in HEP has typically been performed by what we will call the
``cut method.'' The physicist constructs a number of interesting variables,
either representing kinematic observables or a statistical match between data
and predictions. These variables are then individually required to be greater
and/or less than some specified value. This increases the purity of the data
sample at the expense of selection efficiency. Less commonly, techniques which
combine these physics variables in pre-determined ways are included. \Gp makes
no pre-supposition on the final form of a solution.

The charm photoproduction experiment FOCUS is an upgraded version of
FNAL-E687~\cite{Frabetti:1992au} which collected data using the Wideband photon
beamline during the 1996--1997 Fermilab fixed-target run.  The FOCUS experiment
utilizes a forward multiparticle spectrometer to study charmed particles
produced by the interaction of high energy photons  ($\langle E \rangle \approx
180~\gev$) with a segmented BeO target.  Charged particles are tracked within
the spectrometer by two silicon microvertex detector systems. One system is
interleaved with the target segments~\cite{Link:2002zg}; the other is
downstream of the target region. These detectors provide excellent separation
of the production and decay vertices.  Further downstream, charged particles
are tracked and momentum analyzed by a system of five multiwire proportional
chambers~\cite{Link:2001dj} and two dipole magnets with opposite polarity.
Three multicell threshold \cer detectors are used to discriminate among
electrons, pions, kaons, and protons~\cite{Link:2001pg}. The experiment also
contains a complement of hadronic and electromagnetic calorimeters and muon
detectors.

To apply the \gp technique to FOCUS charm data, we begin by identifying a
number of variables which may be useful for separating charm from non-charm. We
also identify variables which may separate the decays of interest from other
charm decays. Many of the variables selected are those we use as cuts in
traditional analyses, but since \gp combines variables not in \emph{cuts}, but
in an algorithmic fashion, we also include a number of variables that may be weak
indicators of charm or the specific decay in question. For instance, we have
included variables that attempt to provide an indicator of opposite (or away)
side charm. 

\subsection{Definition of Fitness}

For the \dcs decay \kpipidcsd, we are looking for very small signals. The fitness of
a tree, as returned to the GPF, describes how small the error on a measurement
of $\text{BR}(\kpipidcsd)/\text{BR}(\kpipi)$  will be.

We would like to maximize the expected significance of the possible signal,
$S_\text{DCS}/\sqrt{S_\text{DCS}+B_\text{DCS}}$, where $S_\text{DCS}$ and 
$B_\text{DCS}$ are the \dcs signal and background, respectively.\footnote{In
the example presented in this paper the number of events is such that the
Gaussian approximation is always valid.}  However, with a small number of
signal events and a very large search space, this method is almost guaranteed
to optimize on fluctuations. Instead, one may choose to \emph{predict} the
number of signal events, $S_\text{DCS}$, from the behavior of the corresponding
\cf decay mode.  In our test of this method on \kpipidcsd, we use the PDG
branching ratio to estimate our sensitivity. Assuming equal selection
efficiencies, $S_\text{CF}$ (the signal yield in the \cf mode) is  proportional
to the predicted number of \dcs events. $S_\text{CF}$ is determined by a fit to
the \cf invariant mass distribution and $B_\text{DCS}$ is determined by a
linear fit over the background range (excluding the signal region) in the \dcs
invariant mass distribution.

However, because we are still optimizing based on $B$ from the \dcs mass plot,
we must be concerned about causing downward fluctuations in $B_\text{DCS}$
which would \emph{appear} to improve our significance. (To a much lesser
extent, we must be concerned about upwardly fluctuating the \cf signal.) We
address this in two ways. First, we apply a penalty to the fitness (or
significance) based on the size of the tree; \ie, we attempt to ensure that any
increase in the tree size is making a significant contribution to reducing the
background (or increasing the signal), not just changing a few
events.\footnote{This has a side effect of preferring the smaller of two
algorithmically identical solutions. It is often noted in \gp that significant
portions of the program may be ``worthless'' in loose analogy to the large
amounts of DNA (introns) in organisms that do not represent genes (this is some
99\% of DNA in humans).} Second, we optimize the significance only on 
even-numbered events. We can then look at the odd-numbered events to assess any
biases. 

Because the \gp framework is set up to minimize, rather than maximize,
the fitness and because we want to enhance differences between small changes in
significance, we minimize the quantity
\begin{equation}
\frac{S_\text{pred} + B_\text{DCS}}{S_\text{CF}^2} \times 10,000 \times (1 +
0.005\times\text{\# of nodes}) 
\eqnlabel{nosignal_fitness}
\end{equation}
where
\begin{equation}
 S_\text{pred} = S_\text{CF} \times
 \frac{\text{BR}(\kpipidcsd)}{\text{BR}(\kpipi)} \,.
\eqnlabel{spred_def}
\end{equation}
The relative branching ratio is taken from the PDG. This penalty factor of
0.5\% per node is arbitrary and is included to encourage the production of
simpler trees. This is more fully explained in \secref{bias}. Squaring
$\sqrt{S+B}/S$ emphasizes differences between trees and multiplying by 10,000
ensures that the average fitness is $\mathcal{O}(1)$.

Finally, we make two additional constraints. We require that at least 500 
events, in \cf and \dcs modes combined, are selected.  We also require that the
\cf signal be observed at the $6\sigma$ level. Both of these requirements
ensure that a good \cf signal exists. Trees which fail these requirements are
assigned a fitness of 100, a very poor fitness.

While it would be preferable to maximize $S/\sqrt{S+B}$ on signal and background
from \mc, there are a number of problems with this. First, we know that our \mc does
not reproduce backgrounds particularly well since non-charm backgrounds are not
simulated. Second, we aren't sure the \mc reproduces all the kinematic and
other parameters of our charm signals correctly, and we certainly don't know
that the \mc would reproduce them correctly in all the combinations the GPF
could create. We can make limited tests of this agreement on the final
product, as discussed in \secref{dplus_mc_comp}. However, because the two \cf
and \dcs decay modes under study are so similar in particle ID and kinematics,
we believe we are justified in assuming that they will behave nearly
identically under various programs suggested by the GPF\@.\footnote{If close agreement
between simulation and data were deemed important, a term to ensure this could
be added to the fitness definition.} 
%Another difference
%between \mc and data is different resonant substructures (and efficiencies) in
%data and the simulation; the effect of this difference is explored in
%\secref{systematics}.

\subsection{Functions and Variables}

We supply a wide variety of functions and variables to the GPF which can be
used to construct trees.  The constructed tree is evaluated for every 
combination. Events for which the tree returns a value greater than zero are
kept. Fits are made to determine $S_\text{CF}$ and $B_\text{DCS}$.

\textbf{Mathematical Functions and Operators:} The first group of functions are
standard mathematical (algebraic and trigonometric) functions and operators. Every
function must be valid for all possible input values, so division by zero and
square roots of negative numbers must be allowed. These mathematical functions and
the protections used are shown in \tabref{operator_math}.

\begin{table}
\centering
\caption{Mathematical functions and operators. $f(n)$ is the sigmoid function commonly used
in neural networks.}
\tablabel{operator_math}
\begin{tabular}{|c|l|}
\hline
Operator & Description \\ \hline
$+$      &                \\
$-$      &                \\
$\times$ &                \\
$/$      & Divide by 0 $\to$ 1 \\ 
$x^y$    & $x$ is 1$^\text{st}$ argument, $y$ is second        \\
$\log$   & $\log\left|x\right|$, $\log 0 = 0$ \\ 
$\sin$   &  \\ 
$\cos$   &  \\ 
$\sqrt{\vphantom{\left|x\right|}\ }$   & $\sqrt{\left|x\right|}$ \\ 
$\text{neg}$   & negative of $x$ \\ 
sign     & returns $(-1,0,+1)$ \\ 
$f(n)$   & $ 1/\left(1+e^{-n}\right)$ \\ 
$\max$   & Maximum of two values \\ 
$\min$   & Minimum of two values \\ 

\hline
\end{tabular}
\end{table}

\textbf{Boolean Operators:} We also use a number of Boolean operators. Our
Boolean operators must take all floating point values as inputs. We define any
number greater than zero as ``true'' and any other number as ``false.'' Boolean
operators return 1 for ``true'' and 0 for ``false.''  The \routine{IF} operator
is special; it returns the value of its second operand if the value of its
first operand is true; otherwise it returns zero (or false). We also use the
comparison operator, $<\!=\!>$, as defined in the Perl programming
language~\cite[pg.\ 87]{Wall:perl_prog}. This operator returns $-1$ if the
first value is less than the second, $+1$ if the opposite is true, and $0$ if
the two values are equal. The Boolean operators are listed in
\tabref{operator_bool}. 

\begin{table}
\centering
\caption{Boolean operators and the comparison operator.}
\tablabel{operator_bool}
\begin{tabular}{|c|l|}
\hline
Operator & Description \\ \hline
$>$      &                \\
$<$      &                \\
AND &  \\ 
OR  &  \\ 
NOT &  \\ 
XOR &  True if one and only one operand is true\\ 
IF  &  2$^\text{nd}$ value if 1$^\text{st}$ true, 0 if false\\
$<\!=\!>$    & $> \to +1$, $< \to -1$, $= \to 0$     \\
 \hline
\end{tabular}
\end{table}

A large number of variables, or terminals, are also supplied. These can be
classified into several groups: 1) vertexing and tracking, 2) kinematic
variables, 3) particle ID, 4) opposite side tagging, and 5) constants. All
variables have been redefined as dimensionless quantities.

\textbf{Vertexing Variables:} The vertexing and tracking variables are 
mostly those used in traditional analyses~\cite{Frabetti:1992au}:  \lsig,
isolation cuts, vertex CLs and isolations, etc. The tracking
\chisq variables which are calculated by the wire chamber tracking
code~\cite{Link:2001dj} are not generally used in analyses. The vertexing and tracking variables are shown in
\tabref{variable_vertex}.

\begin{table}
\centering
\caption{Vertexing and tracking variables. The tracking \chisq variables differ
by species.}
\tablabel{variable_vertex}
\begin{tabular}{|c|c|l|}
\hline
Variable & Units & Description \\ \hline
$\ell$         & cm      & Distance between production and decay vertices      \\
$\sigma_\ell$  & cm    & Error on $\ell$               \\
\lsig          & ---    & Significance of separation between vertices               \\
ISO1, ISO2 &  --- & Isolation of production and decay vertices\\ 
OoM & $\sigma$ & Decay vertex out of material \\ 
POT & $\sigma$ & Production vertex out of target  \\ 
CLP, CLS& --- & CL of production and decay vertices  \\ 
$\sigma_t$ & ps & Lifetime resolution                   \\
$\chisq_{K}$, $\chisq_{\pi_1}$, $\chisq_{\pi_2}$ & --- &  Track \chisq for
$\kaon$, $\pion$, $\pion$  \\
\hline
\end{tabular}
\end{table}

\textbf{Kinematic Variables:} Of the kinematic variables, most are familiar in
HEP analyses. The most prominent exception is $\Sigma p_T^2$ which is the sum
of the squares of the daughter momenta perpendicular to the \emph{charmed parent}
direction.  When large, this variable means the invariant mass of the parent
particle is generated by large opening angles rather than highly asymmetric
momenta of the daughter particles.  In this category, we also include the binary
variables NoTS and TS which represent running before and after the installation
of a second silicon strip detector~\cite{Link:2002zg}. The kinematic variables
are shown in \tabref{variable_kin}.

\begin{table}
\centering
\caption{Kinematic variables.}
\tablabel{variable_kin}
\begin{tabular}{|c|c|l|}
\hline
Variable & Units & Description \\ \hline
$\#\tau$ & --- &  Lifetime/mean lifetime \\  
$p$ & \gevc & Charm momentum \\  
$p_T$ &  \gevc & $p$ transverse to beam\\  
$\Sigma p_T^2$ & $\text{GeV}^2/c^2$ &  Sum of daughter $p_T^2$ (see text) \\  
$m_\text{err}$ & \mevcc &  Error on reconstructed mass \\  
TS   & 0, 1 &  Early, late running \\  
NoTS & 0, 1 &  Opposite of TS \\  
\hline
\end{tabular}
\end{table}

\textbf{Particle Identification:} For particle ID, we use the standard FOCUS
\cer variables~\cite{Link:2001pg} for identifying protons, kaons, and pions. We
also include Boolean values from the silicon strip tracking code for each of
the tracks being consistent with an electron (zero-angle) and the maximum CL
that one of the decay tracks is a muon. The particle ID input variables are
shown in \tabref{variable_pid}.

\begin{table}
\centering
\caption{Particle ID variables.}
\tablabel{variable_pid}
\begin{tabular}{|c|c|l|}
\hline
Variable & Units & Description \\ \hline
$\Delta \pion \kaon_1$ & --- &  \kaon not \pion \\  
$\picon{}_1$         & --- &  \pion consistency, first pion \\  
$\picon{}_2$         & --- &  \pion consistency, second pion \\  
$\mu_\text{max}$       & --- &  Maximum \muon CL of all tracks\\  
$K_e$, $\pi_{e1}$, $\pi_{e2}$ & 0/1 (True/False) &  Electron consistency from silicon
tracker \\  
\hline
\end{tabular}
\end{table}

\textbf{Opposite Side Tagging:} Since \gp needn't cut on variables we have
investigated some possible methods for opposite (or away) side tagging. 
Any cut on these tagging methods is too inefficient to be of any use, but
combined with other variables, variables representing the probability of
the presence of an opposite side decay may be useful. 
Three such
variables were formed and investigated. The first attempts to construct charm
vertices from tracks which are not part of the production \emph{or} decay
vertices. The best vertex is chosen based on its confidence level. The second
general method of opposite side tagging is to look for kaons or muons from
charm decays that are not part of the decay vertex. (Here, tracks in the
production vertex are included since often the production and opposite side
charm vertices merge.) We determine the confidence level of the best muon of
the correct charge and the kaonicity of the best kaon of the correct charge not
in the decay vertex. The variables for opposite side charm tagging are shown in
\tabref{variable_osc}. 

\begin{table}
\centering
\caption{Opposite side tagging variables.}
\tablabel{variable_osc}
\begin{tabular}{|c|c|l|}
\hline
Variable & Units & Description \\ \hline
$\text{CL}_\text{opp}$ & --- &  CL of highest vertex opposite \\  
$\dwpik_\text{opp}$ & --- &  Highest kaonicity not in secondary  \\  
$\text{CL}\mu_\text{opp}$ & --- &  Highest muon CL not in secondary\\  
\hline
\end{tabular}
\end{table}

\textbf{Constants:} In addition to these variables, we also supply a number
of constants. We explicitly include 0 (false) and 1 (true) and allow the GPF
to pick real constants on the interval $(-2,+2)$ and integer
constants on the interval $(-10,+10)$.

The optimization in this example uses 21 operators or functions and 34
terminals (variables or constants).

\subsection{Anatomy of a Run}

Once we have defined all of our functions and terminals and defined the
fitness, we are ready to start the GPF. To recap, the steps taken by the GPF
are:
\begin{enumerate}
  \item{Generate a population of individual programs to be tested.}
  \item{Loop over and calculate the fitness of each of these programs.}
  \begin{enumerate}
    \item{Loop over each physics event.}
    \item{Keep events where the tree evaluates to $>0$, discard other events.}
    \item{For surviving events, fit the \cf signal and \dcs background.}
    \item{Return the fitness calculated according to \eqnref{nosignal_fitness}.}
  \end{enumerate}    
  \item{When all programs of a generation have been tested, create another
  generation by selecting programs for breeding according to the selection
  method. Apply breeding operators such as cross-over, mutation, and
  reproduction.}
  \item{Every few generations, exchange the best individuals among ``islands.''}
  \item{Continue, repeating steps 2--4 until the desired number of generations
        is reached.}
\end{enumerate}    

\section{Selecting Genetic Programming Parameters}

There are a large number of parameters that can be chosen within the GPF, such as
numbers of individuals, selection probabilities, and breeding probabilities.
Each of these can affect the evolution rate of the model and affect the
probability that the initial individuals don't  have enough diversity to reliably find a
good minimum. It should be
emphasized, though, that the final result, given enough time, should not be
affected by these choices (assuming sufficient diversity). The default parameters
for the studies presented in this section are given in \tabref{study_gp_param}. 

\begin{table}
\centering
\caption{Default \gp parameters for studies.}
\tablabel{study_gp_param}
\begin{tabular}{|l|r|}
\hline
Parameter & Value \\ \hline\hline
Generations & 6 \\
Programs/sub-population & 1000  \\  
Sub-populations & 20 \\
Exchange interval & 2 generations \\
Number exchanged & 5 \\
\hline  
Selection method & Fitness-over-select \\
Cross-over probability   & 0.85 \\  
Reproduction probability & 0.10  \\  
Mutation probability     & 0.05  \\  \hline
Generation method & Half grow, half full \\
Full depths   & 2--6 \\  
Maximum depth   & 17 \\  
  \hline
\end{tabular}
\end{table}

In monitoring our test runs, we look at the fitness and size of each
individual. We only consider individuals which have a fitness less than about 3
times worse than the fitness of a single-node tree which
selects all events. This excludes individuals where no
events were selected and others where much more signal than background was
removed. We then look at average and best fitness as a function of generation
while we vary various parameters of the \gp run.

In \figref{doubling-evolution} we show the effects of various ways of doubling
the total number of programs examined by the \gp run. We begin with a run on 20
sub-populations with 1000 individuals per sub-population and lasting 6
generations. We then double each of these quantities. One can see that either
doubling the sub-population size or doubling  the number of sub-populations
gives about the same best and average fitness as the original case. However,
doubling the number of generations has a significant effect on the average and
best fitness of the population. After 12 generations (plus the initial
generation ``0''), evolution is still clearly occurring.

\begin{figure}
  \begin{center}
  \includegraphics[width=12.5cm,bb=35 390 570 740]{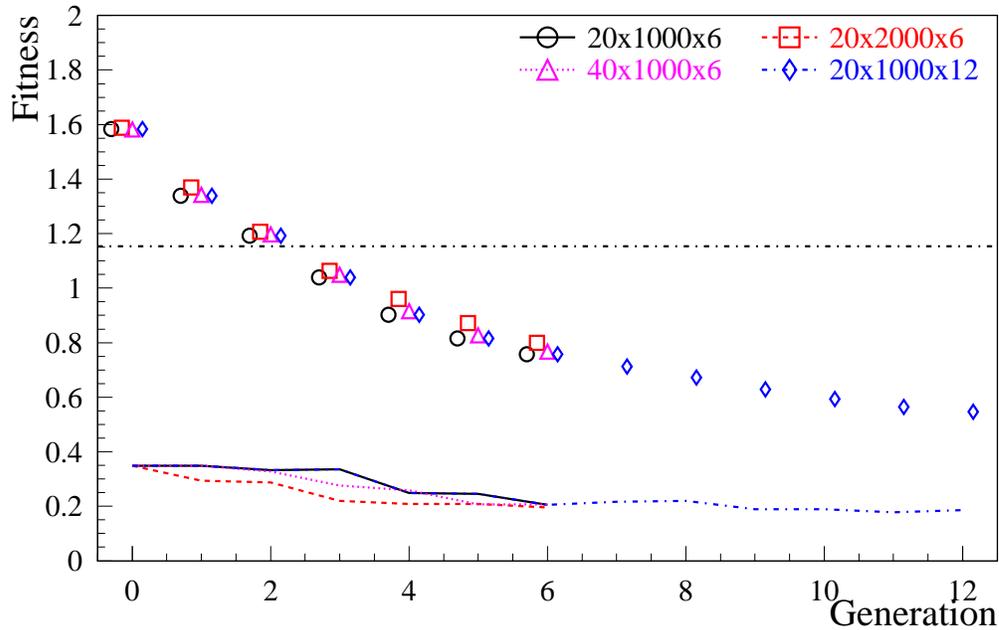}
  \end{center}
  \caption[Plots of fitness vs. generation for increased statistics]{Plots of
  fitness vs. generation for increased statistics. The points show the average
  fitness for a generation and the curves show the fitness of the best
  individual. The straight line shows the fitness of the initial sample. The
  circles and solid line show the initial conditions: 20 sub-populations, 1000
  individuals per sub-population, and 6 generations. The squares and dashed
  line show the effect of doubling the individuals per sub-population to 2000.
  The triangles and dotted line show the effect of doubling the sub-populations
  to 40. The diamonds and dotted-dashed line show the effect of doubling the
  number of generations to 12.}
  \figlabel{doubling-evolution}
\end{figure}

In addition to changing the number of individuals evaluated during the \gp run
as discussed above, there are a number of parameters in the \gp framework that
can be adjusted. These may change the rate or ultimate end-point of the
evolution.

In \figref{parameter-evolution} we show the effect of changing some of the
basic parameters of the \gp framework on the evolution of the population. These
plots show the effect over 12 generations of changing the selection method, the
number of programs exchanged during the migration phase, and the size of the
initial programs generated.

\begin{figure}
  \begin{center}
  \includegraphics[width=12.5cm,bb=35 390 570 740]{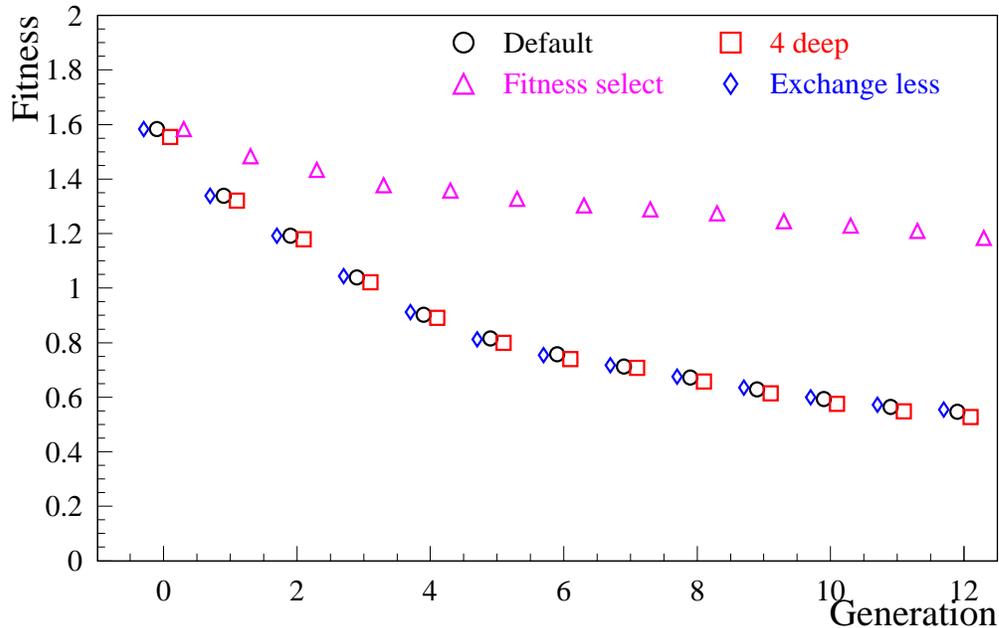}
  \end{center}
  \caption[Plots of fitness vs. generation for GP parameters]{Plots of fitness
  vs. generation for different GP parameters. The points show the average
  fitness for a generation.  The circles show the default settings. The squares
  show the effect of changing the ``full'' part of the initial population
  generation to produce trees with depths of 2--4 (rather than the default
  2--6).  The diamonds show the effect of exchanging 2 rather than 5
  individuals per process in the migration phase. The triangles show the effect
  of using ``fitness'' rather than ``fitness over-select'' selection. (The
  fitness over-select method is used in all other plots.)}
  \figlabel{parameter-evolution}
\end{figure}

Exchanging 2 rather than 5 individuals from each process every 2
generations results in no change in the evolution rate.
(For the standard case where we have 20 CPUs each with a population of 1000,
exchanging 5 individuals from each of 20 CPUs means that 10\% of the resulting
sub-population is in common with each other sub-population.)

Changing the selection method used has a dramatic effect on the evolution.
Changing from our default ``fitness-over-select'' method to the more common
``fitness'' method results in a much more slowly evolving population. Recall
that in the standard fitness selection, the probability of an individual being
selected to breed is directly proportional to the fraction of the total fitness
of the population accounted for by the individual. In the fitness-over-select
method, individuals in the population are segregated into two groups based on
their fitness. Individuals which breed are selected more often from the more
fit group (which also tends to be smaller). Selection within each group is done
in the same way as for the standard fitness selection method.

Conventional wisdom is that fitness-over-selection can be useful for large populations
(larger than about 500 individuals).  For smaller populations, there are risks
associated with finding local minima.

As discussed in \secref{tree_gen}, half of the trees are generated by the
``full'' generation method. In the default evolution plot shown in
\figref{parameter-evolution}, the beginning depths of these trees are from 2 to
6. So, 10\% of the initial trees are of depth 2, 10\% are of depth 3, etc., up
to  10\% of depth 6. The remaining 50\% are generated by the ``grow'' method. As can be
seen, changing the depth of the full trees so that trees of depth 5 and 6 are
not allowed has little effect on the evolution rate. Other, earlier studies
indicate that this change may positively affect the fitness in early
generations and negatively affect the fitness of the later generations.

\figref{mutation-evolution} shows the effect of changing the mutation
probabilities. Increasing the mutation probability from 5\% to 10\% (at the
expense of the crossover probability) does very little to change the evolution
rate. Reducing the mutation probability to 2.5\% or even 0\% has a similarly
small effect. However, at higher mutation rates, we worry about the effects of
too high a mutation rate on later generations where stability should be
increasing. 

\begin{figure}
  \begin{center}
  \includegraphics[width=12.5cm,bb=35 390 570 740]{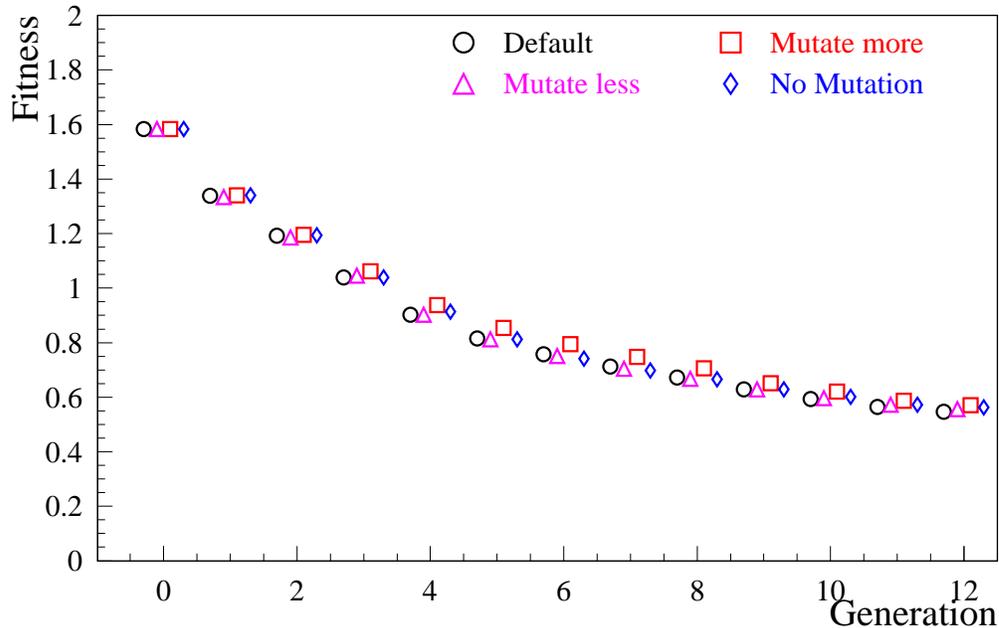}
  \end{center}
  \caption[Plots of fitness vs. generation for mutation probability]{Plots of
  fitness vs. generation for different mutation rates. The points 
  show the average fitness for a generation.  The 
  circles and show the default settings.  The squares show the
  effect of doubling the mutation probability from 5\% to 10\%, the 
  triangles show the effect of halving the mutation probability from 5\% to
  2.5\%, and the diamonds show the effect of eliminating mutation.}
  \figlabel{mutation-evolution}
\end{figure}

\figref{functions-evolution} shows the effect of reducing the number of
functions used in the search. In the first case, we removed a number of the
algebraic, trigonometric, logical, and arithmetic operators. Operators such as
$\log$, $x^y$, $\sin$, $\cos$, XOR, and others were not allowed to be used.
This reduces the diversity of the program set, but may allow a minimum to be
found faster if those operators are not necessary. In another trial, we removed
variables not normally used in FOCUS analyses, such as opposite side tagging,
track CLs, muon and electron mis-ID for hadrons, etc. In a final trial we
removed both non-standard variables and questionable functions. The assumption
in all three cases is the same: if the functions or variables added to the
minimal case do not positively contribute to the overall fitness, slower
evolution and possibly a worse final solution should be observed since the
useful parameter space is less well covered. That this is not observed in the
average case (which is less susceptible to fluctuations) suggests
that the ``extra'' functions and variables are useful and ultimately better
solutions may be found by including them.

\begin{figure}
  \begin{center}
  \includegraphics[width=12.5cm,bb=35 390 570 740]{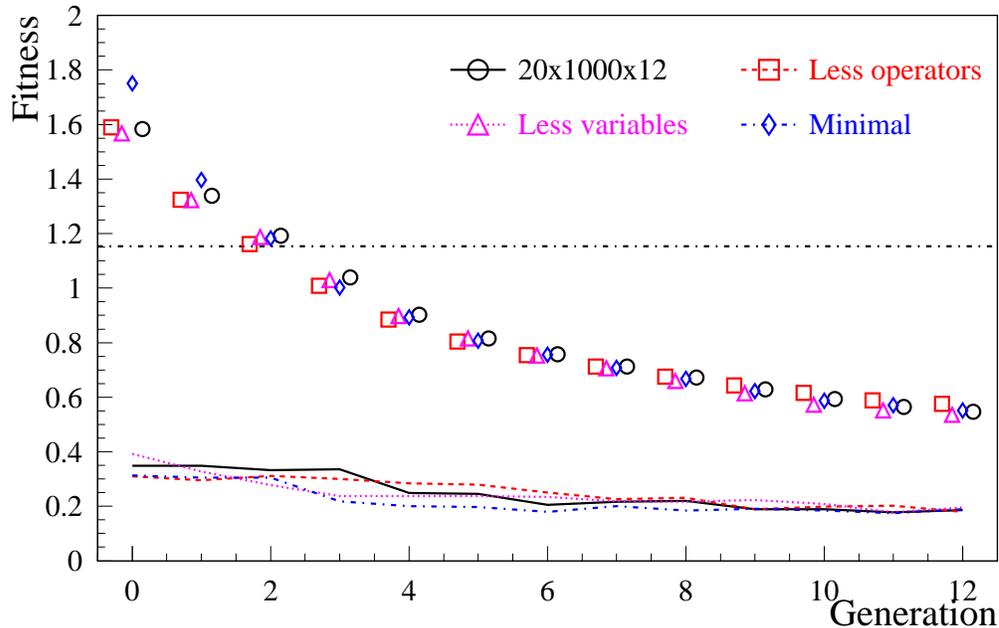}
  \end{center}
  \caption[Plots of fitness vs. generation for reduced functions]{Plots of
  average and best fitness vs. generation for sets of reduced functions. For
  ``Less operators,'' (squares and dashed line) the set of arithmetic,
  algebraic, and logical operators was reduced to a minimal set. For ``Less
  Variables'' (triangles and dotted line) only ``standard'' analysis variables
  were included. For ``Minimal'' (diamonds and dashed-dotted line) both
  reductions were employed.}
  \figlabel{functions-evolution}
\end{figure}

For all the studies detailed so far, we used a data sample with a relatively
strict cut of $\lsig>15$. This is a very powerful cut and is an easy way to
obtain a small, highly enriched sample of \dmeson mesons, speeding up the
studies. However, the goal of our use of the \gp method is to try to discover
ways \emph{around} making such tight cuts. \figref{initial-evolution} shows the
effect of changing the cut on the initial data sample to $\lsig>10$. One sees
that while initially the best individuals are not as pure as the $\lsig>15$
sample, parity is quickly reached. It's reasonable to assume that in later
generations, the $\lsig>10$ selection becomes more effective as events with
$10<\lsig<15$ (about 7.5\% of the total in the 12th generation) are included
which have other indications of being \kpipi decays. (The plots of the average
never reach the same level, but this is to be expected as the initial purity is
not as good and many programs, even at later stages, do not perform any event
selection. Because of this effect, comparing the averages of the two cases is
not instructive.)

\begin{figure}
  \begin{center}
  \includegraphics[width=12.5cm,bb=35 390 570 740]{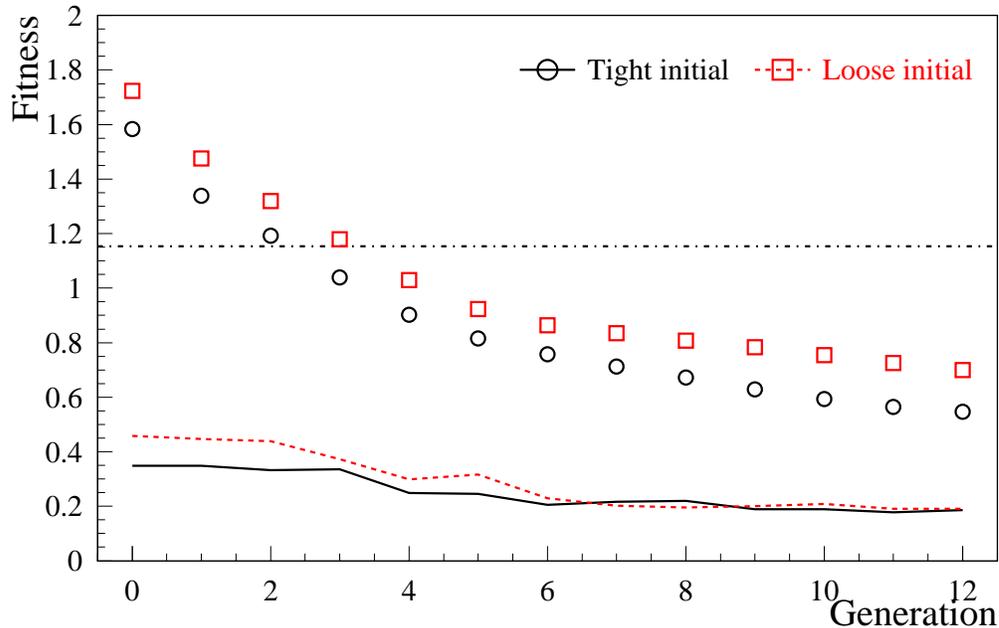}
  \end{center}
  \caption[Plots of fitness vs. generation for different initial data samples]{Plots of fitness vs. generation for
  different initial data samples. The default sample (circles and solid line)
  has a cut of $\lsig>15$ while the ``Loose'' sample has a cut of $\lsig>10$.}
  \figlabel{initial-evolution}
\end{figure}

%\clearpage % Temporary?

\section{Testing \GP on \kpipidcsd}\seclabel{dplus}

Having explored various settings in the GPF to determine the way to obtain the
best evolution of programs, we now investigate the ultimate performance and
accuracy of this method to measure the relative branching ratio
$\text{BR}(\kpipidcsd)/\text{BR}(\kpipi)$.

Recall from \eqnref{nosignal_fitness} that the GPF is attempting to
minimize the quantity
\begin{equation}
\frac{B_\text{DCS}+S_\text{pred}}{S_\text{CF}^2} \times 10,000 \times (1 +
0.005\times\text{\# of nodes}) \, ,
\eqnlabel{signal_fitness}
\end{equation}
where $B_\text{DCS}$ is a fit to the background excluding the $\pm 2\sigma$
signal window and $S_\text{pred}$ is the expected \dcs yield determined from
the PDG value for the \kpipidcsd relative branching ratio~\cite{pdg:pdg2002}
and the \cf signal ($S_\text{CF}$). We exclude from the fit the  expected
signal region in the \dcs decay channel to avoid inflating $B_\text{DCS}$ (or
alternatively allowing the GPF to learn how to eliminate the \dcs
signal).\footnote{Since the background is fit to a straight line, including the
\dcs signal would increase the apparent level of the background,
\emph{reducing} the calculated sensitivity to the \dcs decay.}

To start with a data sample of manageable size, certain
requirements must be enforced on the data supplied to the GPF. For the final
runs looking for \kpipidcsd, these requirements are:
\begin{itemize}
 \item{\lsig $>10$}
 \item{CLS $>0.01$}
 \item{\picon $>-8.0$ for both pions}
 \item{\dwpik $>1.0$ for kaon}
 \item{$1.75~\gevcc < \text{Mass} <1.95~\gevcc $}
\end{itemize} 
These requirements are applied to both \kpipi and  \kpipidcsd  candidates.

The initial data sample is shown in \figref{dp-data-orig}. The \cf signal
dominates, and the level of \dcs candidates (the linear histogram) is higher than
the \cf background. The \cf fit finds $253\,180\pm660$ events.
  
\begin{figure}
\centering
 \includegraphics[width=11cm]{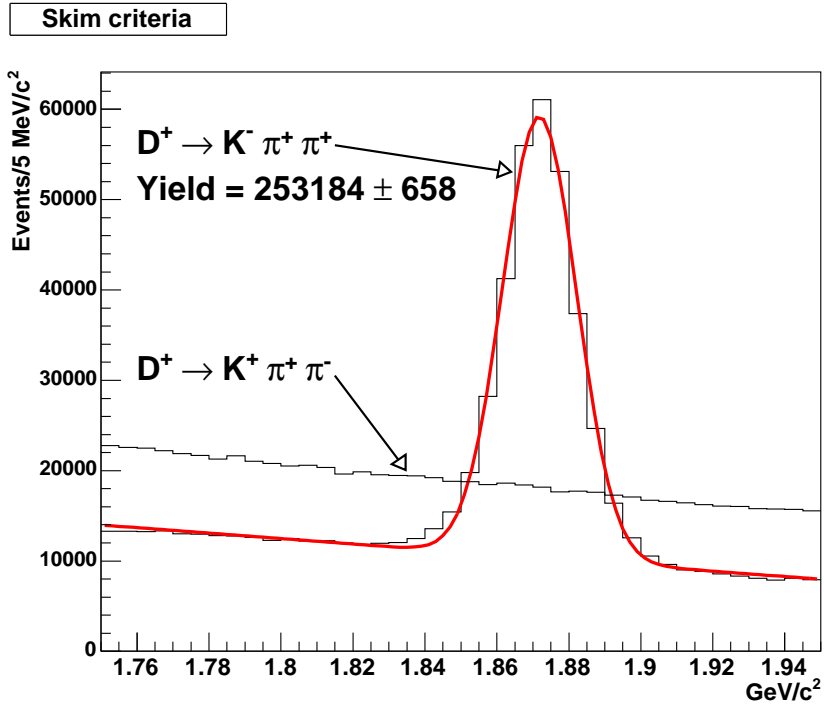}
\caption{The initial \kpipi and \kpipidcsd candidate distributions (the cuts are described in the
text). The linear \dcs invariant mass distribution is
superimposed on the \cf distribution. (No signal is visible.)}
\figlabel{dp-data-orig}
\end{figure}

We select the \kpipidcsd and \kpipi events using the genetic programming
technique with the parameters listed in \tabref{anal_gp_param}. These parameters
are similar to those used in our earlier studies, but we increase the programs
per generation and the number of generations.

\begin{table}
\centering
\caption{\Gp parameters for \kpipidcsd optimization.}
\tablabel{anal_gp_param}
\begin{tabular}{|l|r|}
\hline
Parameter & Value \\ \hline\hline
Generations & 40 \\
Programs/sub-population & 1500  \\  
Sub-populations & 20 \\
Exchange interval & 2 generations \\
Number exchanged & 5 \\
\hline  
Selection method & Fitness-over-select \\
Cross-over probability   & 0.85 \\  
Reproduction probability & 0.10  \\  
Mutation probability     & 0.05  \\  \hline
Generation method & Half grow, half full \\
Full depths   & 2--6 \\  
Maximum depth   & 17 \\  
  \hline
\end{tabular}
\end{table}

\figref{dp-data} shows the data after selection by the GPF for the \cf and \dcs
decay modes. The GPF ran for 40 generations. A \dcs signal is now clearly
visible.  $62\,440\pm260$ (or about 25\%) of the original \cf events remain and
$466\pm36$ \dcs events are now visible. The \dcs background has been reduced by
a factor greater than 150. All fits to \dcs signals use the mass and
resolution determined from the \cf signals. The free parameters are the yield
and the background shape and level.

\begin{figure}
\centering
%\subfigure[\cf signal and fit ($62\,440\pm260$ events)]{
% \includegraphics[width=6.6cm,bb=0 0 567 384,clip]{dp-cf.eps}
%}
%\subfigure[\Dcs signal and fit ($466\pm36$ events)]{
% \includegraphics[width=6.6cm]{dp-dcs.eps}
%}
 \includegraphics[width=13.8cm]{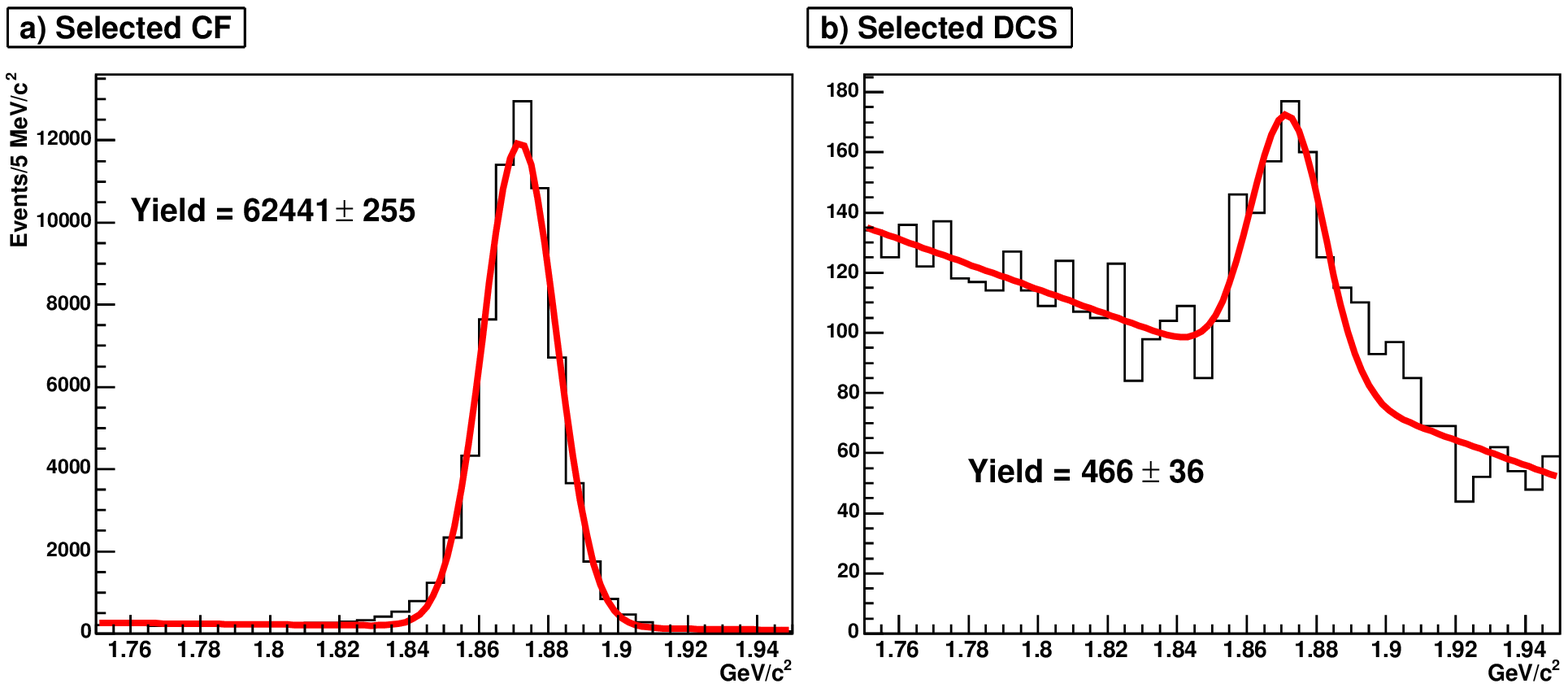}
\caption{\kpipi (a) and \kpipidcsd (b) signals selected by \gp. We find
$62\,440\pm260$  and $466\pm36$ events, respectively.}
\figlabel{dp-data}
\end{figure}

The evolution of the individuals in the \gp is shown in \figref{dp-evolution}.
In addition to the variables plotted before, the
average size for each generation is also plotted. We can see that the average
size reaches a minimum at the 4th generation, nearly plateaus between the 20th
and 30th generations, and then begins increasing again and is still increasing
at the 40th generation.

\begin{figure}
\centering
 \includegraphics[width=13.9cm]{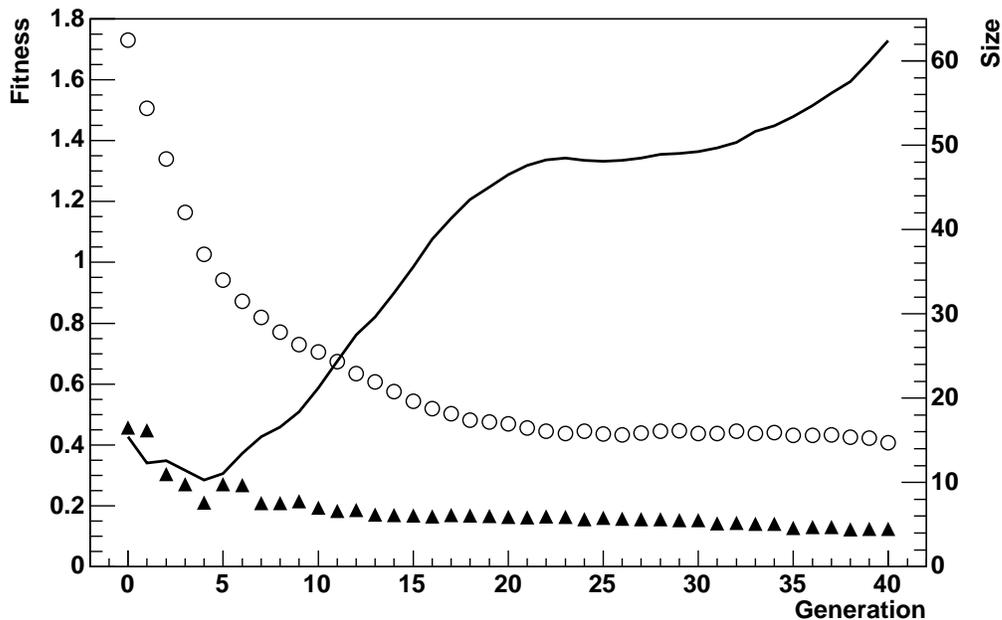}
\caption[Evolution of \kpipi programs.]{Evolution of \kpipi programs. The open
circles show the average fitness of the population as a function of generation.
The triangles show the fitness of the best individual. The solid line shows the
average size of the individuals as a function of the generation.}
\figlabel{dp-evolution}
\end{figure}

In \figref{dp-evolution}, the average and best fitnesses seems to stabilize,
but in \figref{dp-blowup} an enlargement of the best fitness for later
generations is seen. From this plot, it is apparent that evolution is still
occurring at the 40th generation and new, better trees are still being found.

\begin{figure}
\centering
 \includegraphics[width=13.9cm]{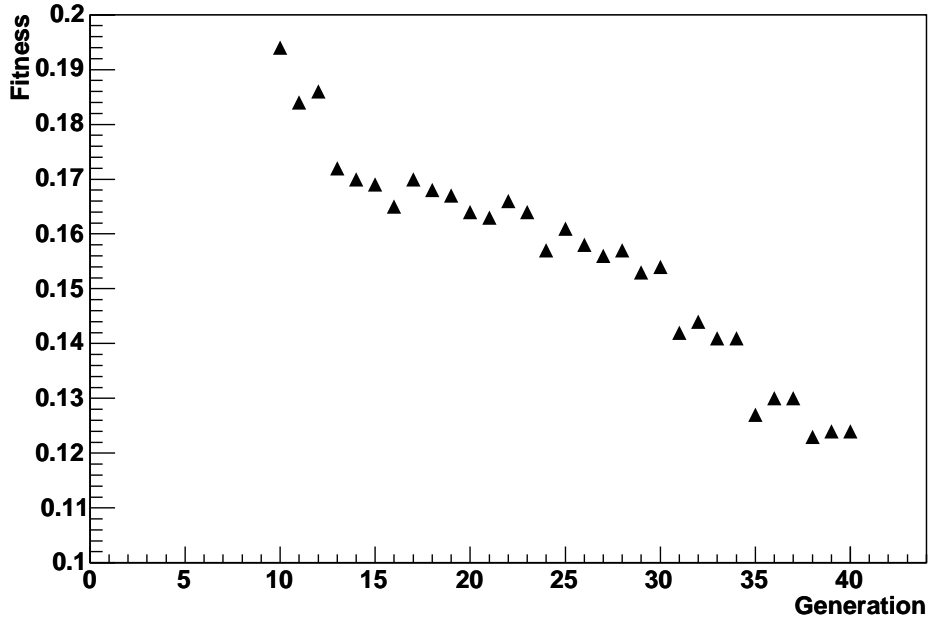}
\caption{Expanded view of evolution of \kpipi programs. The triangles show the 
fitness of the best individuals for later generations, the same data as in
\figref{dp-evolution}, but on an expanded scale.}
\figlabel{dp-blowup}
\end{figure}

\subsection{Example trees from various generations}\seclabel{example_trees}

In \figref{gen0-examples} we show four of the most fit trees from the initial
generation (numbered 0); no evolution has taken place at this point. It is
interesting to note several things. First, the best trees are all rather small,
although the average tree in this generation is over 15 nodes in size. Second,
the most fit tree in this generation (a) is recognizable to any FOCUS
collaborator: it requires that the interaction vertex point be in the target
and that the decay vertex point be located outside of target material. The
second most fit tree (b) is algorithmically identical to the first, but has a
slightly worse fitness because it is considerably larger than the first tree.
Tree (c) is nearly identical in functionality to (a) and (b) but actually does
a slightly better job of separating events than (a) or (b). Its fitness is
slightly worse because of its larger size (12 nodes vs.\ 5 and 10 nodes
respectively).

\begin{figure}
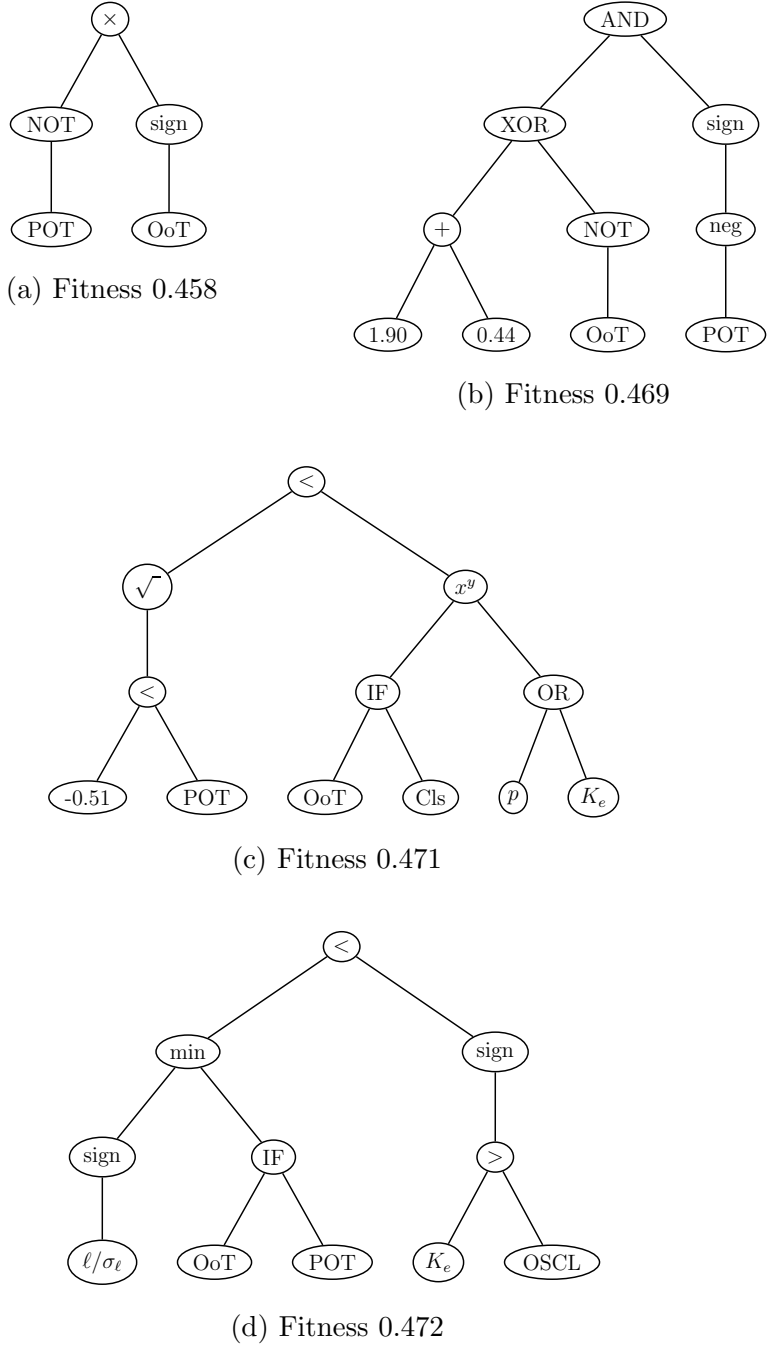
 \centering \subfigure[Fitness 0.458]{
\scalebox{0.7}{\makebox[8cm]{\input{trees/dp_gen0_458.tex}}} }
\subfigure[Fitness 0.469]{ \scalebox{0.7}{\input{trees/dp_gen0_469.tex}} }
\subfigure[Fitness 0.471]{ \scalebox{0.7}{\input{trees/dp_gen0_471.tex}} }
\subfigure[Fitness 0.472]{ \scalebox{0.7}{\input{trees/dp_gen0_472.tex}} }
\caption{The four most fit trees from generation 0 of a \kpipidcsd run.}
\figlabel{gen0-examples} \end{figure}

The four best individuals from generation 2 are shown in
\figref{gen2-examples}. A few observations are in order. First, the most fit
tree (a) doesn't include either of the elements we saw in the best trees in
generation 0 (OoT and POT), but this tree is significantly more fit than any
tree in generation 0. The second best tree (b) does include these elements,
mixed in with others. Also, one can see that the fourth most fit tree (d) is
quite large (44 nodes) compared to the other best trees in the generation.

\begin{figure}
\centering
\subfigure[Fitness 0.305]{
 \scalebox{0.6}{\input{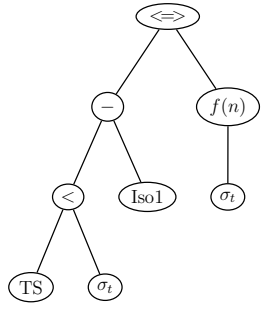}}
}
\subfigure[Fitness 0.309]{
 \scalebox{0.5}{\input{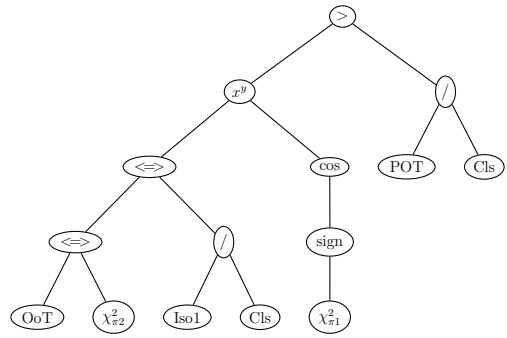}}
}
\subfigure[Fitness 0.383]{
 \scalebox{0.6}{\input{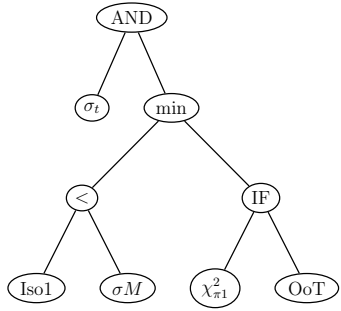}}
}
\subfigure[Fitness 0.397]{
 \scalebox{0.5}{\input{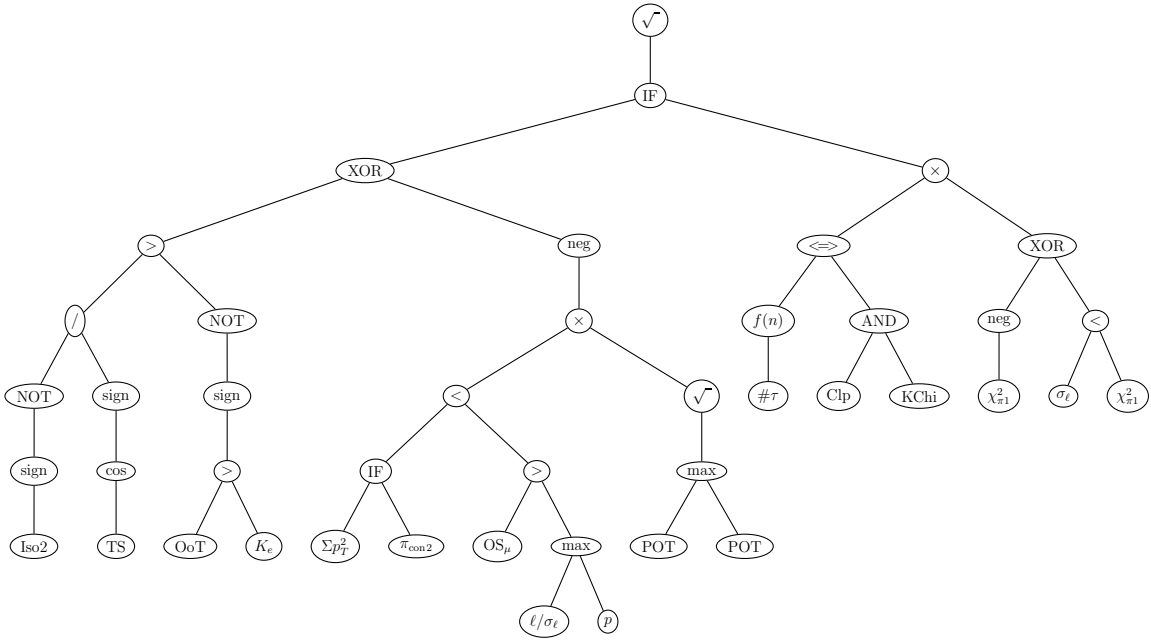}}
}
\caption{The four most fit trees from generation 2 of a \kpipidcsd run.}
\figlabel{gen2-examples}
\end{figure}

As shown in \figref{dp-evolution}, the average size tree of the population
reaches a minimum in generation 4. One possible interpretation of this behavior
is that at this point, the GPF has determined the parts of successful
solutions. Going forward, it is assembling these parts into more complex
solutions. The three best solutions from generation four are shown in
\figref{gen4-examples}.  The best solution from generation four (a) is very
similar to the best solution from generation two, but with the two end nodes
replaced with new end nodes. The second and third best solutions are clearly
related, with just one small difference: the second best tree (b), with the
sub-branch: $\text{OS}_\mu <\!=\!> \text{OoT}$ only lets about 250 events with
OoT $\le 0$ through the filter while the third best tree (c) allows about 9300
such events through.

\begin{figure}
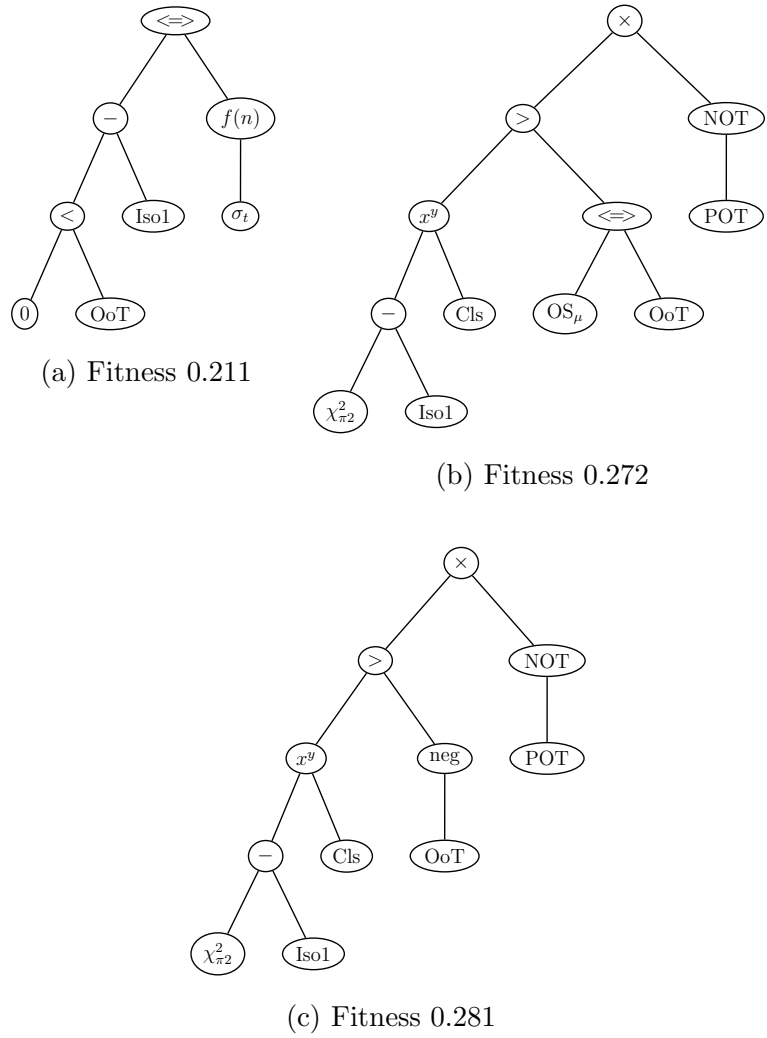

\centering
\subfigure[Fitness 0.211]{
 \scalebox{0.65}{\input{trees/dp_gen4_211.tex}}
}
\subfigure[Fitness 0.272]{
 \scalebox{0.65}{\input{trees/dp_gen4_272.tex}}
}
\subfigure[Fitness 0.281]{
 \scalebox{0.65}{\input{trees/dp_gen4_281.tex}}
}
%\subfigure[Fitness 0.328]{
% \scalebox{0.65}{\input{trees/dp_gen4_328.tex}}
%}
\caption{The three most fit trees from generation 4 of a \kpipidcsd run.}
\figlabel{gen4-examples}
\end{figure}

In \figref{gen10-examples}, we can see that tree (a) and tree (b) have some
pieces in common, but are generally quite different. They are also about equally good at
separating background from signal, but tree (b) has the larger fitness due to
its larger size.

\begin{figure}
\centering
\subfigure[Fitness 0.194]{
 \scalebox{0.65}{\input{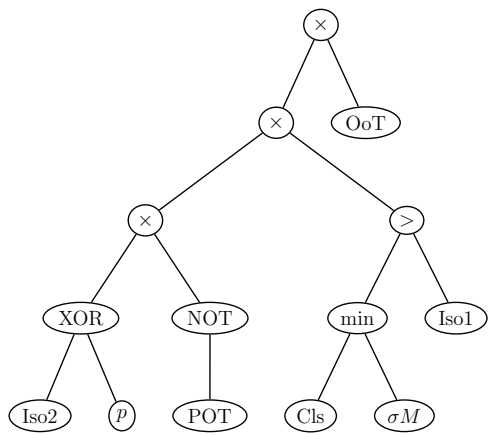}}
}
\subfigure[Fitness 0.198]{
 \scalebox{0.65}{\input{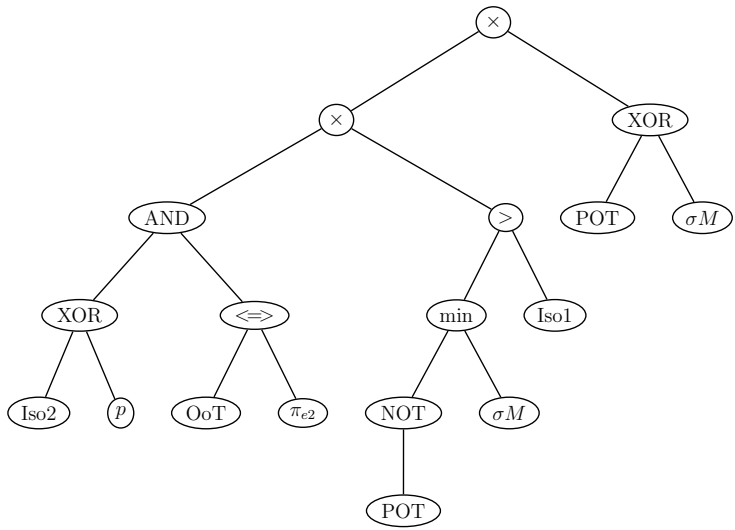}}
}
\caption{The two most fit trees from generation 10 of a \kpipidcsd run.}
\figlabel{gen10-examples}
\end{figure}

\begin{figure}
\centering
\subfigure[Most fit tree: fitness 0.1234]{
 \scalebox{0.54}{\input{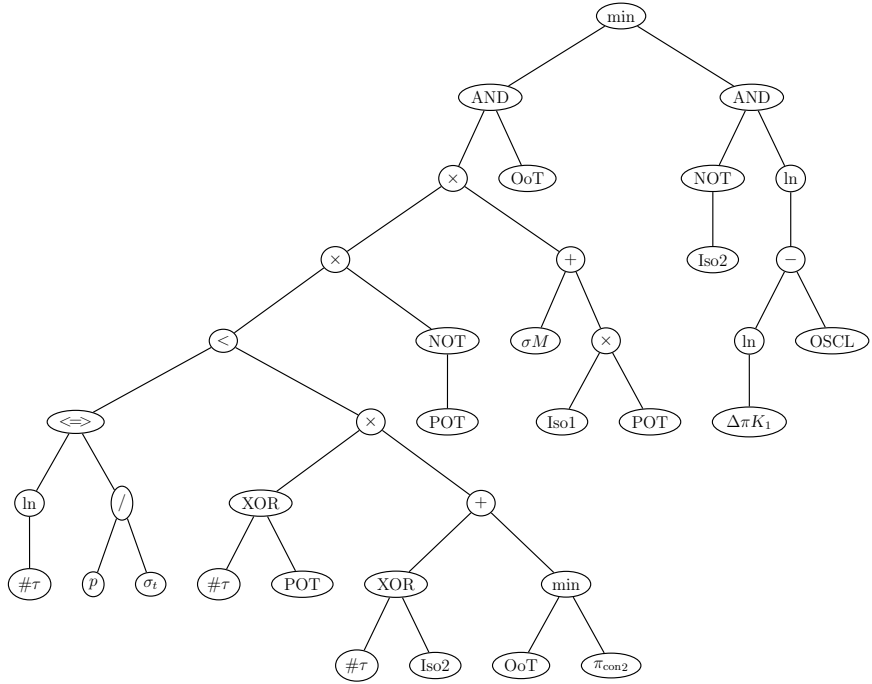}}
}
\subfigure[Third most fit tree: fitness 0.1285]{
 \scalebox{0.54}{\input{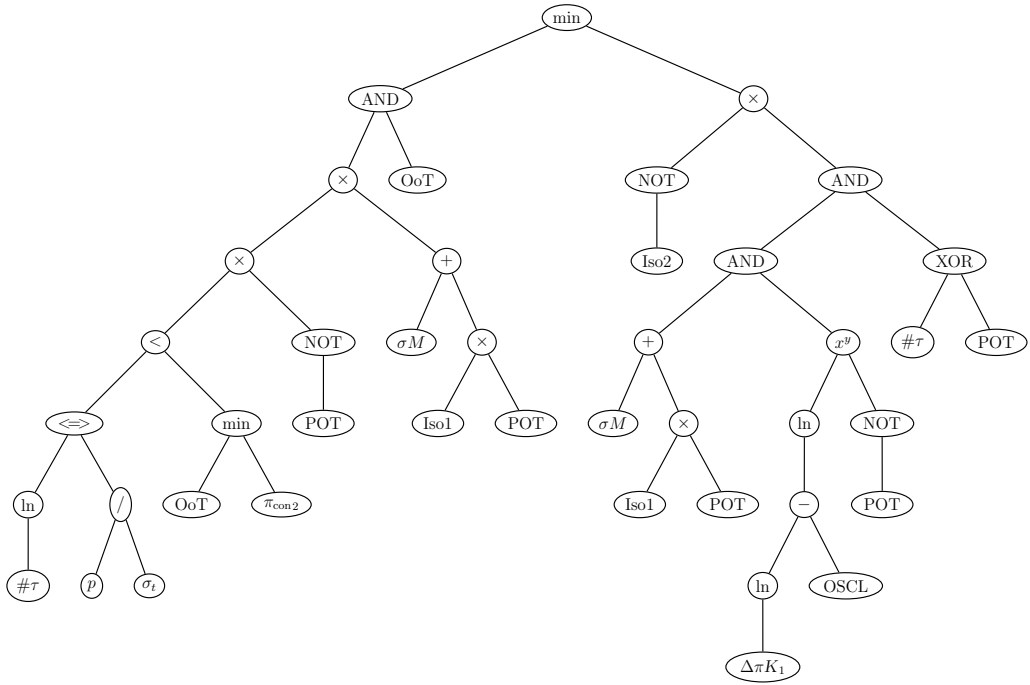}}
}
\caption{The two most fit trees from generation 40 of a \kpipidcsd run. (The
second most fit tree is a duplicate of the most fit.}
\figlabel{final-kpipi}
\end{figure}

\clearpage % Temporary?
\section{Systematic Error Studies}\seclabel{systematics}

As with any analysis method, the \gp technique may introduce sources of
systematic error into the analysis. In our case, there are several ways of evaluating this
possible error. 

\subsection{Efficiency corrections and comparison with traditional methods}\seclabel{dplus_mc_comp}

When we optimize, we assume that the efficiencies of the filter trees are
identical for \cf and \dcs decays. While we can't test this on the data, since
we lack an observable \dcs signal in the initial sample, we can test this assumption with
our \mc.  \mc simulations of $\sim 4\times10^6$ events in each of the two decay
modes show that the efficiencies for the \dcs and \cf modes are nearly
consistent. These results are summarized in \tabref{dp_mc_calc}.

\begin{table}
\centering
\caption{\mc efficiencies for \kpipi and \kpipidcsd.}
\tablabel{dp_mc_calc}
\begin{tabular}{|l|r|r|}
\hline
Decay mode & Skim efficiency & GPF efficiency \\ \hline
\kpipi     &  $5.76\pm0.01$\% &    $1.434 \pm 0.007$\%            \\
\kpipidcsd &  $5.57\pm0.01$\% &    $1.408 \pm 0.007$\%         \\
\hline
\end{tabular}
\end{table}

In the data we find $466\pm36$ \dcs decays and $62\,440\pm260$ \cf decays. 
Along with the relative efficiencies in \tabref{dp_mc_calc}, this gives a
corrected branching ratio of $(0.76 \pm 0.06)$\%. The PDG value is $(0.75 \pm
0.16)$\%~\cite{pdg:pdg2002}; a recent analysis from FOCUS sets this value at
$(0.65 \pm 0.08 \pm 0.04)$\%~\cite{Link:2004mx}. The data sample used for the
branching ratio measurement is shown in \figref{edera}.\footnote{The Cabibbo
suppressed decay $\dsplus \to \kplus \piminus\piplus$ is also visible in this
plot. We remove this mass region to simplify fitting.} One can see that while
the standard analysis used $189\pm24$ events, the \gp method finds $466\pm36$
with similar signal to noise. However, the standard analysis was not optimized
for $S/\sqrt{S+B}$, so a direct comparison is not possible.

\begin{figure}
 \begin{center}
  \includegraphics[width=8cm,bb=0 0 520 520]{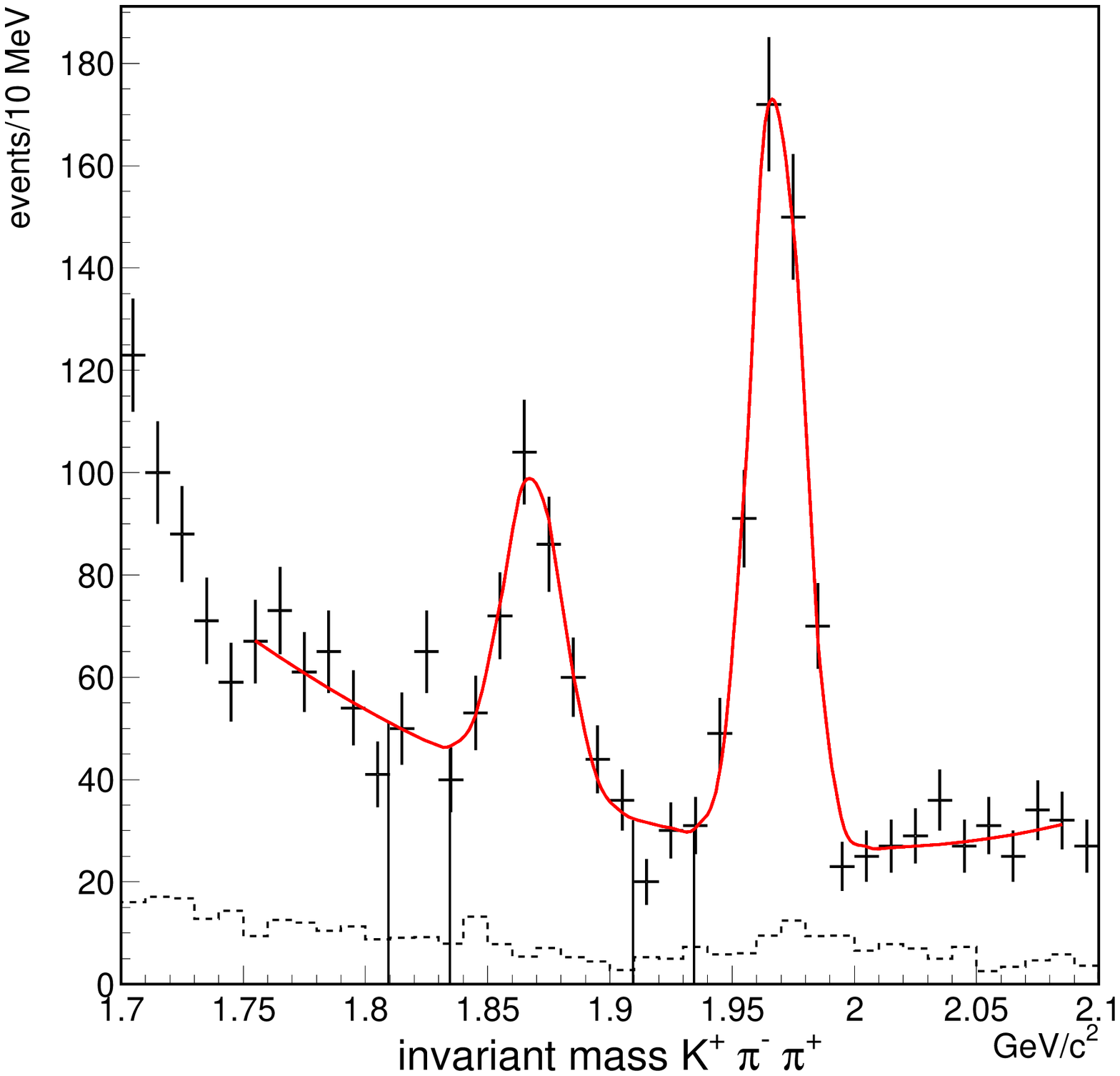}
 \end{center}
 \caption{The invariant mass plot of \kpipidcsd from an earlier FOCUS
          publication~\cite{Link:2004mx}. The Cabibbo suppressed decay $\dsplus
          \to \kplus \piminus \piplus$ is also visible in this plot.}
 \figlabel{edera}
\end{figure}

From this check, it is apparent that \gp can yield correct results, even if we
don't understand exactly how. This study also suggests that \gp can yield
results with greater sensitivity than our standard analysis methods.

\subsection{Efficiency comparisons with \mc}

An additional, more challenging, check against the FOCUS \mc is possible. We
can measure the efficiency of the final tree on the \mc and data. For this
study, we look at the ratios of relative efficiency:
\begin{equation}
\frac{\epsilon_\text{GP-data}}{\epsilon_\text{skim-data}} \, \text{ and }
\frac{\epsilon_\text{GP-MC}}{\epsilon_\text{skim-MC}}
\eqnlabel{xcheck_eff}
\end{equation}
for the \cf decay mode which reduces to 
\begin{equation}
\frac{Y_\text{GP-data}}{Y_\text{skim-data}} \, \text{ and }
\frac{Y_\text{GP-MC}}{Y_\text{skim-MC}}\, ,
\eqnlabel{xcheck_yield}
\end{equation}
where the various $Y$ values are the yields in \mc or data with only the skim
cuts or the skim cuts \emph{and} the tree selection applied. In other words,
the efficiencies of the tree with respect to the skim on both \mc and data
should be the same if the behavior of the tree is completely modeled by the
\mc. A similar study on \dcs decays is not possible since the yield under the
skim cuts is unknown. The results of this study are shown in
\tabref{dp_cf_xcheck}.

\begin{table}
\centering
\caption{\mc and data  yields \kpipi for skim cuts only and skim cuts plus the
final tree. The values in the final column are the ratios shown in
\eqnref{xcheck_eff}.}
\tablabel{dp_cf_xcheck}
\begin{tabular}{|l|c|c|c|}
\hline
Source & Skim yield & GPF yield & Ratio \\ \hline
Data     &  $253\,190\pm 660$ & $62\,440\pm 260$ &    $24.7 \pm 0.1$\%            \\
\mc      &  $236\,000\pm 500$ & $58\,788\pm 245$ &    $24.9 \pm 0.1$\%         \\
\hline
\end{tabular}
\end{table}

In this test, the \gp tree is very well modeled by the FOCUS \mc. This was
unexpected. We know our \mc matches the one-dimensional and some
two-dimensional distributions of  the included variables. But to well model all
trees generated by the GPF, the \mc must correctly model the interrelationships
between variables --- in this case, a match in 11-dimensional space (one for
each terminal in the final tree that is a unique physics variable). Recall that we choose our
measurement so that we don't require that the \gp tree perform the same on \mc
and data, only that the tree perform the same on \cf and \dcs data. From the
very similar efficiencies shown in \tabref{dp_mc_calc}, we have additional
evidence to believe selection by the \gp tree will not be a significant source
of systematic error in analyses with such similar final states. 

Should close agreement between \mc and data be required, the fitness function
can be redefined to enforce agreement.

\subsection{Bias induced by \gp optimization}\seclabel{bias}

Recall that we attempt to avoid bias by assigning a 0.5\% penalty to the
fitness for each node in the tree. But, the size of this penalty is arbitrary
and is chosen to be small to gently encourage the GPF to produce smaller
trees. 

To study the possibility that the \gp optimization is selecting events based on
their specific properties rather than the general properties of all \kpipidcsd
decays, we only optimize on half the events. We can then look at the other half
of the events to discover any major problems. (We optimize on even-numbered
events, so there is no problem with selecting events from one run period over
another, etc.)

In \figref{dp-cf-bias}, we show the \kpipi candidates used and unused in the
optimization. While the used portion has a few more events, the difference is
statistically insignificant. In \figref{dp-dcs-bias}, we show the \kpipidcsd
candidates; while there are more signal events in the ``used'' plot, recall
that this region is masked out. It is more instructive to look at the
background, since this is susceptible to downward fluctuations. ($B$ is
determined from a linear fit with the signal region masked out.) 
One can see that the background in the two plots is nearly identical. As
mentioned above, all fits to \dcs signals use the mass and
resolution determined from the \cf signals.

\begin{figure}
 \begin{center}
  \subfigure[Used candidates]{\includegraphics[width=6.6cm]{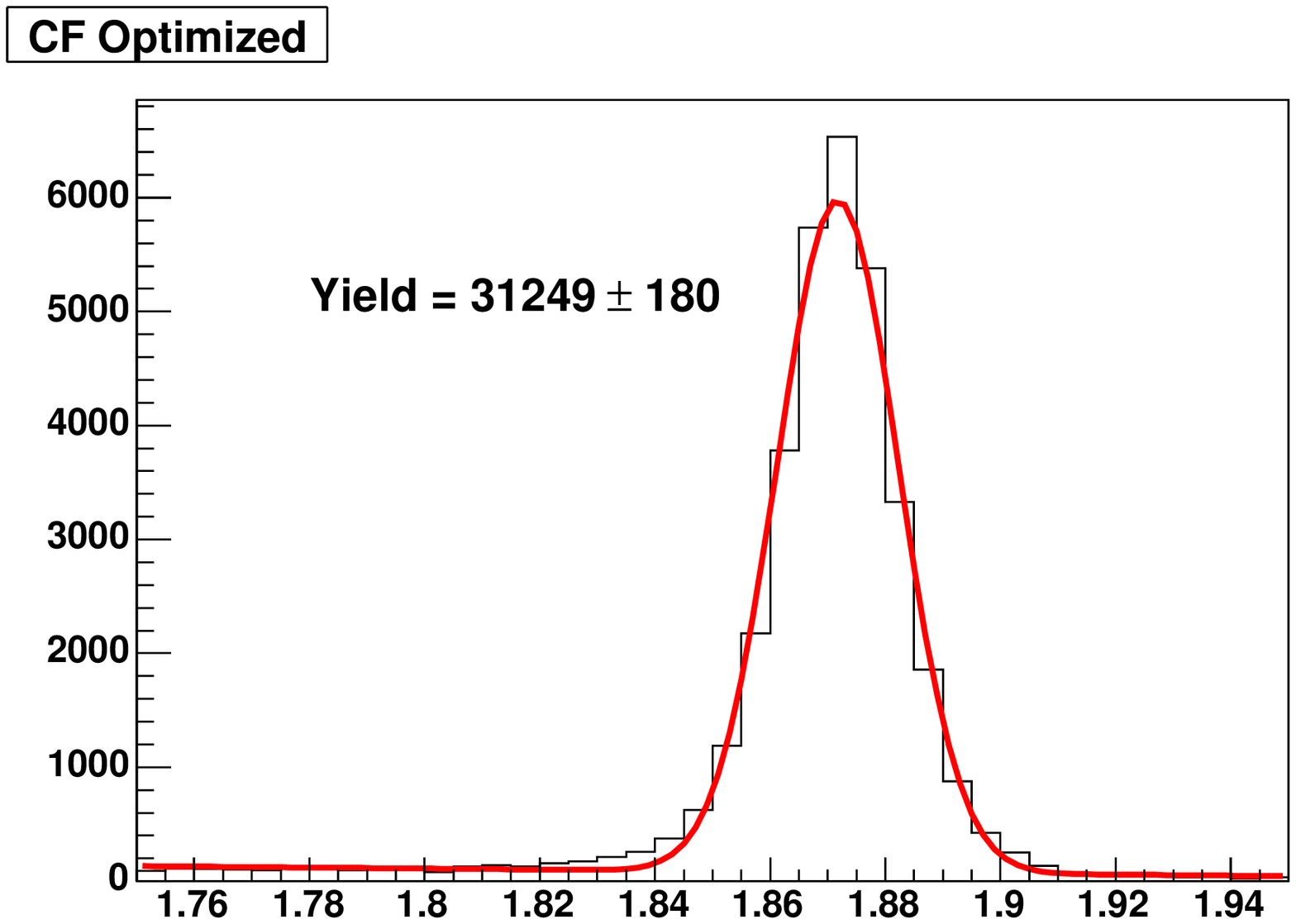}}
  \subfigure[Unused candidates]{\includegraphics[width=6.6cm]{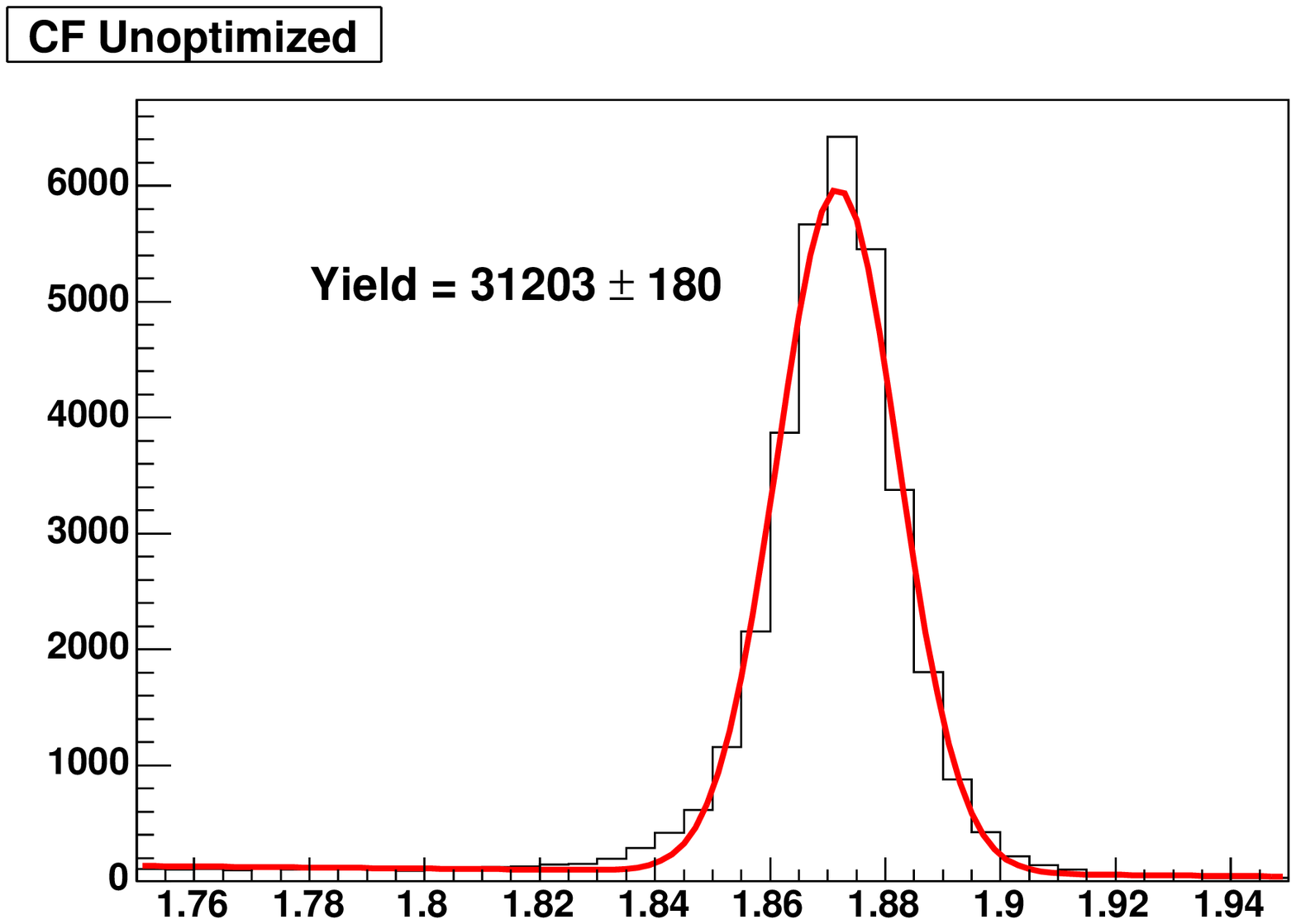}}
 \end{center}
 \caption{The \dplus \cf decay candidates used (a) in the optimization and those not used (b)
 as a cross check. The \dplus yields are $31250\pm180$ and $31200\pm180$ events
 respectively.}
 \figlabel{dp-cf-bias}
\end{figure}

\begin{figure}
 \begin{center}
  \subfigure[Used candidates]{\includegraphics[width=6.6cm]{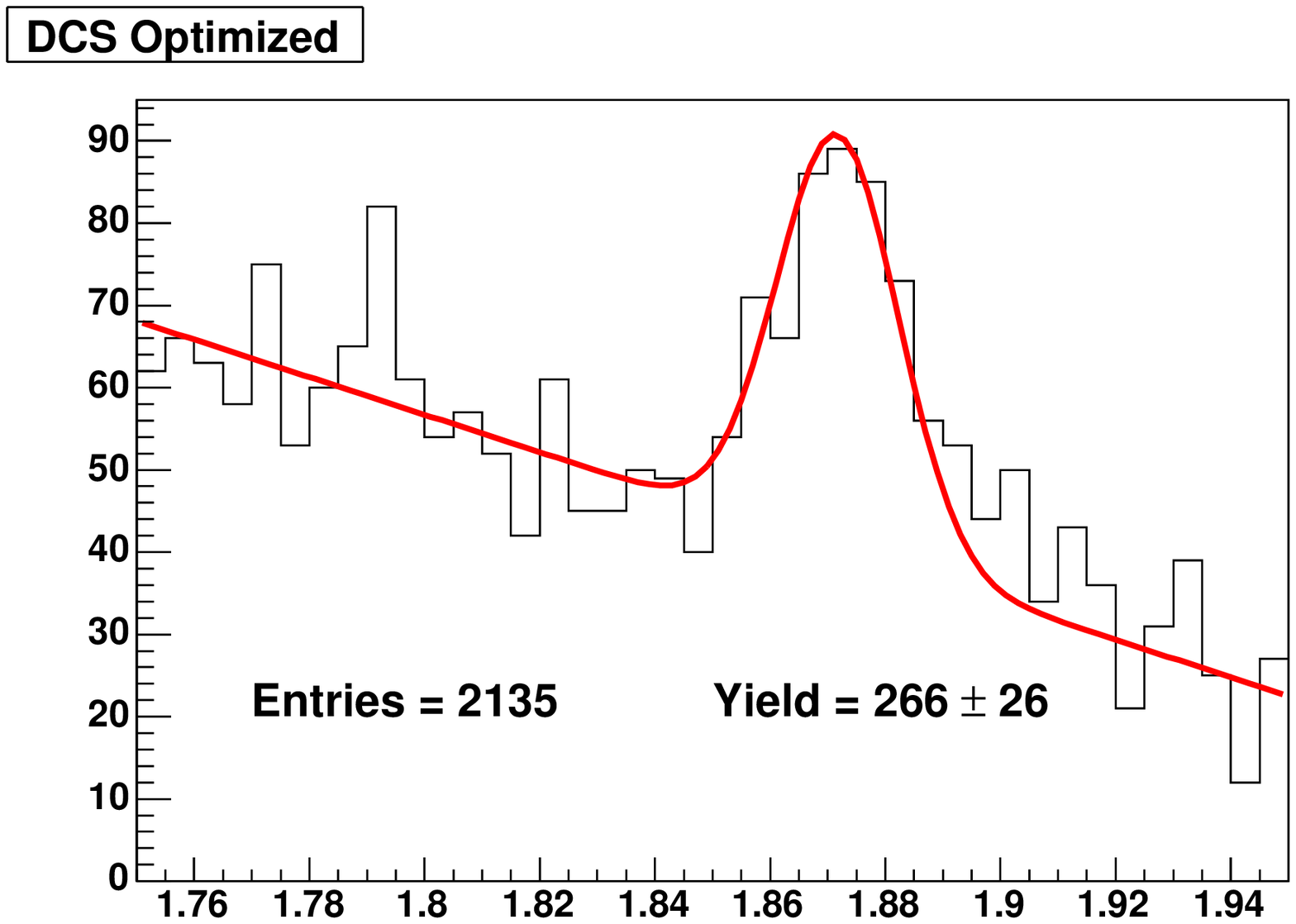}}
  \subfigure[Unused candidates]{\includegraphics[width=6.6cm]{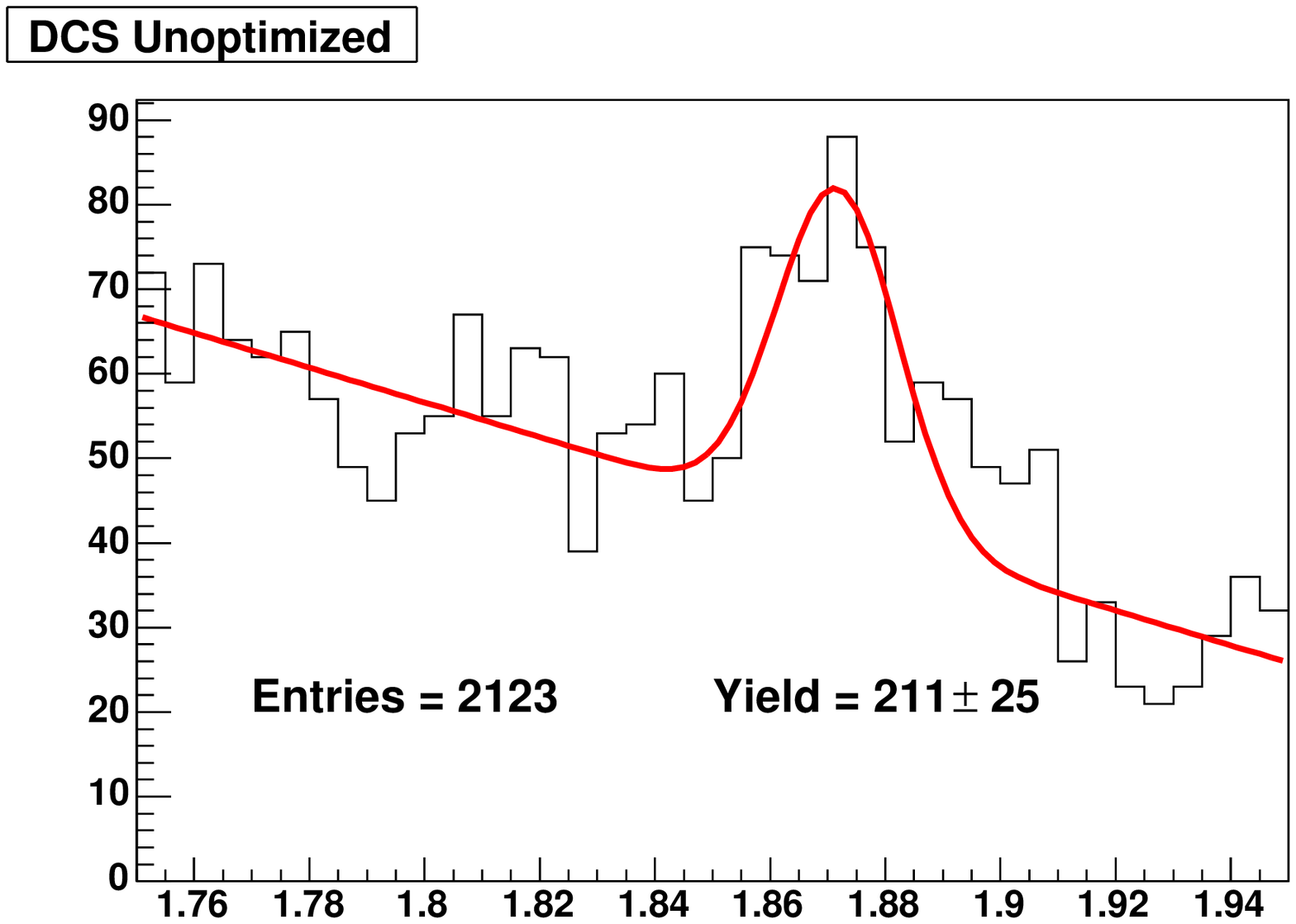}}
 \end{center}
 \caption{The \dplus \dcs decay candidates used (a) in the optimization and those not used (b)
 as a cross check.}
 \figlabel{dp-dcs-bias}
\end{figure}

To further study possible biases induced by the \gp method, analysis of
additional runs optimizing on the other half of the events and/or with a
different random seed to the GPF would be necessary. Even if there is such a
bias, it appears to be small and can easily be incorporated into the systematic
error. 

\subsection{Different Evolutionary Trajectories}

In a traditional cut based analysis, one may choose to make measurements with
several different sets of cuts.  In \gp, one analogous method of assessing such
variations is to change the trajectory of the evolution. We have studied two
ways of doing this. In the first case, we optimize on the odd-numbered rather
than even-numbered events. (The initial trees are identical, but their
estimated fitnesses differ, causing the evolutionary paths to diverge.)  In the
second, we change the random number seed at the beginning of the process. We
also show the results in the ``default'' case, but with only 10 rather than 40
generations of evolution.

We then investigate possible differences in three areas. The first observation
is the overlap in events between the best tree in the ``default'' analysis and
the best five trees from the ``other'' analysis. This is shown in
\figref{overlap}. We define two quantities: ``false positives,'' events which are
selected by a given tree but are rejected by the best tree in the
default analysis, and ``false negatives,'' events which are rejected by a given
tree but are selected by the best tree.\footnote{Note that our terminology
of false positives and negatives treats the best tree from the default case as
``true'' which is not to be confused with which decays are \emph{actually}
\kpipi or \kpipidcsd decays.} We can see that in the default (even-numbered)
case, there are no false positives, but some false negatives. In the other
analyses, up to 20\% of the events are classified differently by different
trees.  

\begin{figure}
 \begin{center}
  \includegraphics[width=13cm]{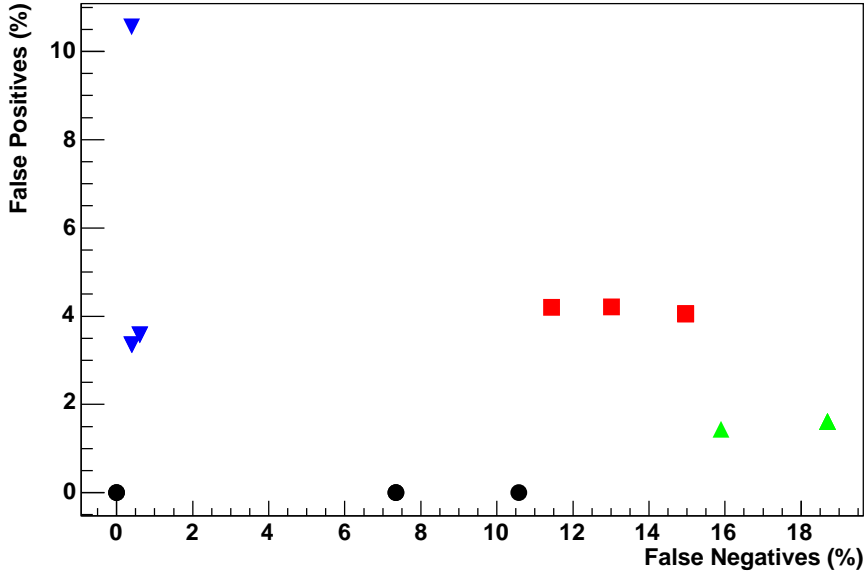}
 \end{center}
 \caption{The percentage of events classified differently by different trees.
 For an explanation of the quantities plotted, see the text. The circles show
 the default optimization, the squares show the results of optimization on
 odd-numbered events, the triangles ($\blacktriangle$) show the effect of a different random number
 seed, and the inverted triangles ($\blacktriangledown$) show the effect of stopping the evolution at
 10 generations. (Though the five best trees are considered, identical or nearly
 identical trees cause the number of visible points to be reduced.)}
 \figlabel{overlap}
\end{figure}

Second,  we observe the \dcs signals obtained from the best tree from each of
these four different evolutionary trajectories. These are shown in
\figref{overlap_histo}.  In all cases, we see a very clear \dcs signal. We can
see that the two trajectories optimizing on even-numbered events for 40
generations [plots a) and c)] obtain similar results. Without a large number of
different runs with different seeds, it is impossible to say if optimizing on
odd vs.\ events would give generally similar results. However, we can certainly
see that we are gaining quite a bit from optimizing for an additional 30
generations. (Compare a) with a $13\sigma$ significance and much better
signal-to-background ratio vs.\ d) with an $11\sigma$ significance.)

\begin{figure}
 \begin{center}
  \includegraphics[width=13cm]{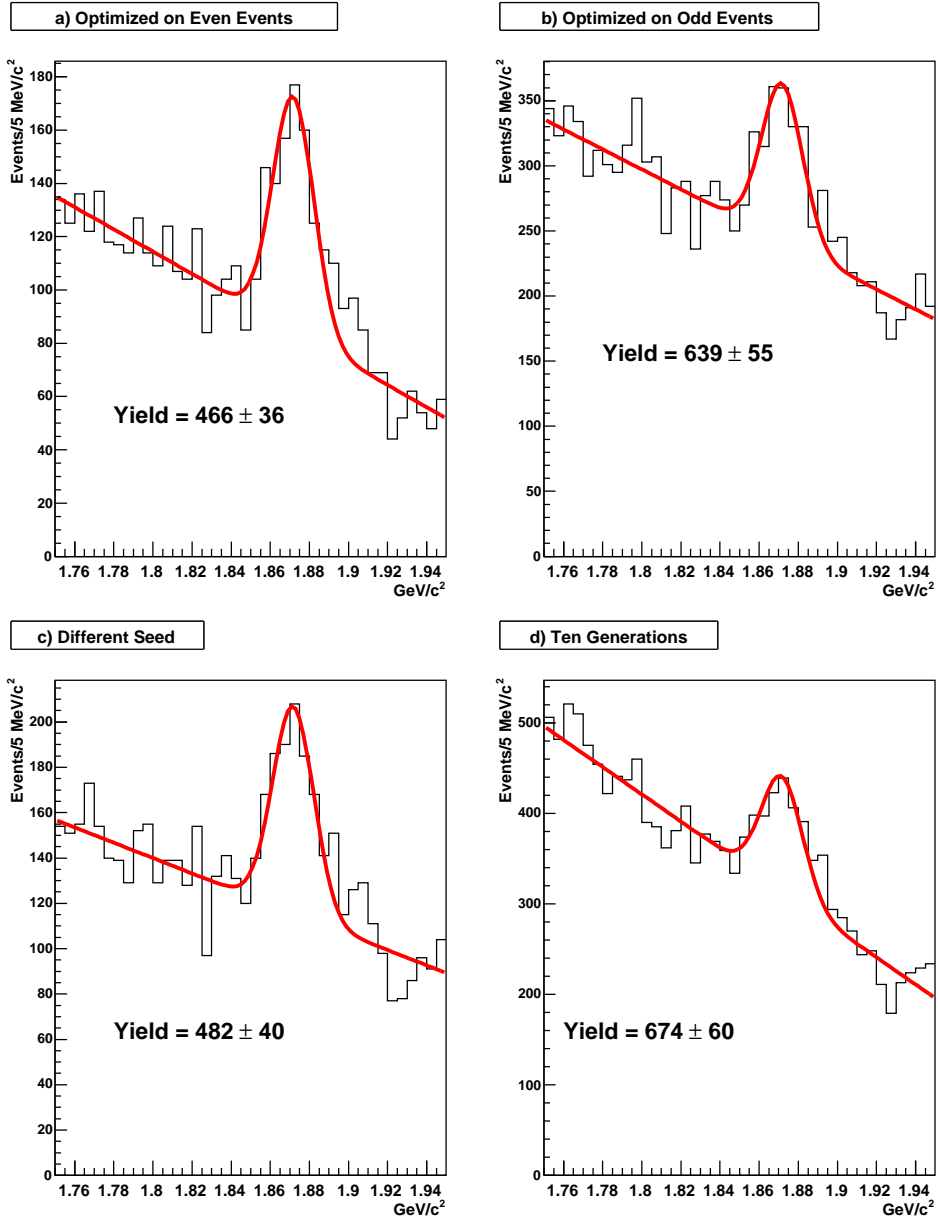}
 \end{center}
 \caption{The \dcs signals obtained from four different evolutionary
 trajectories. a) is the default optimization (with a fitness of 0.123), b)
 shows the results of optimization on odd-numbered events (fitness: 0.148), c)
 shows the effect of a different random number seed (fitness: 0.148), and d)
 shows the effect of stopping the evolution at 10 generations (fitness:
 0.194).}
 \figlabel{overlap_histo}
\end{figure}

Finally, we measure the \dcs branching ratio with the five top trees in each of
the four cases. Because, as shown in \tabref{dp_mc_calc}, the \mc corrections
to the relative efficiency are small, we simply look at the uncorrected
branching ratio, $Y_\text{DCS}/Y_\text{CF}$. The values for these twenty trees
are shown in \figref{top20}. As expected, the five trees within each group give
nearly identical answers, while some variation is evident between groups. Such
a variation, after \mc efficiency corrections and corrections for expected
statistical fluctuations, would form a portion of the systematic error in an
analysis using this technique. 

\begin{figure}
 \begin{center}
  \includegraphics[width=13cm]{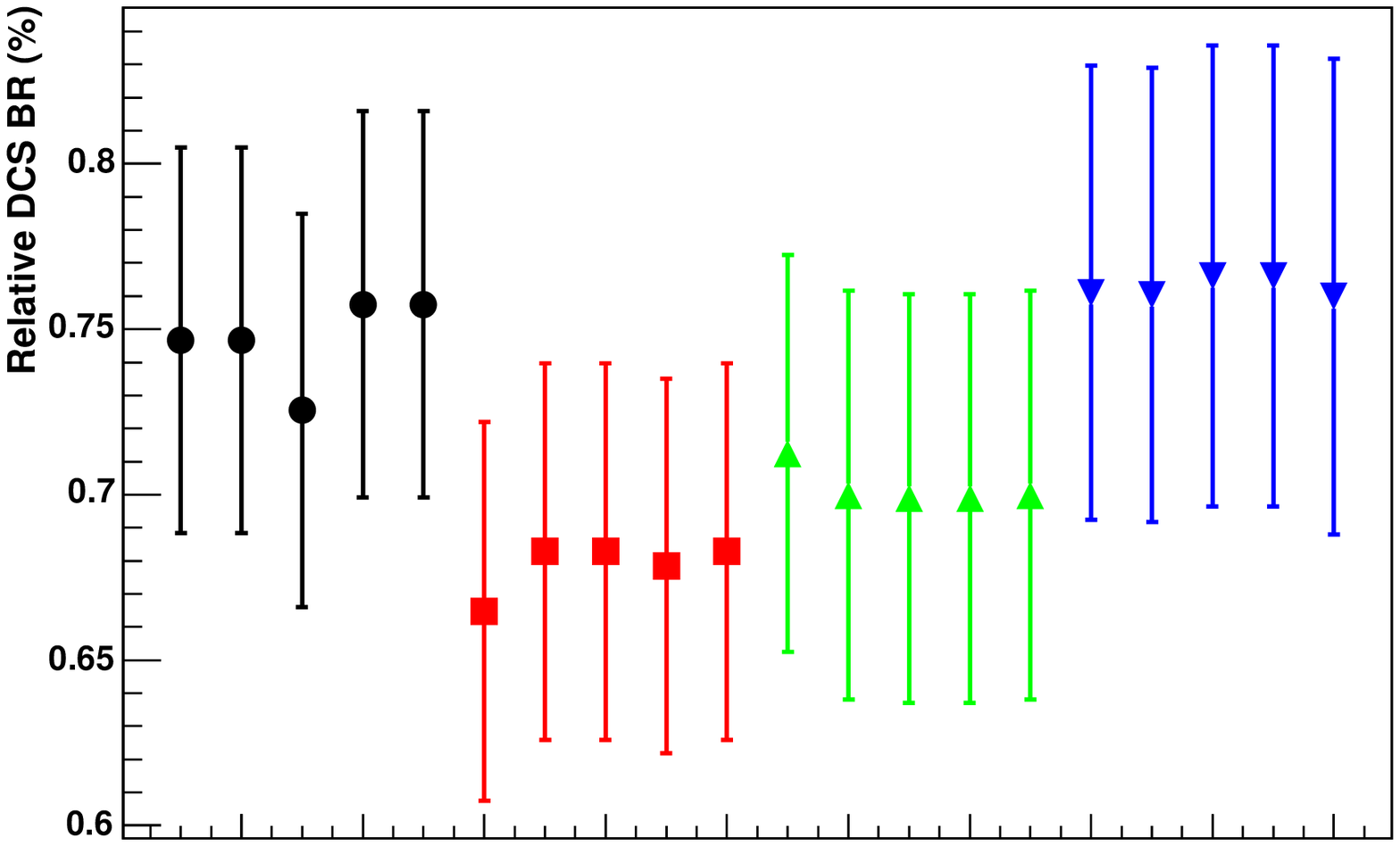}
 \end{center}
 \caption{The branching ratio, uncorrected by \mc, from the five best trees in
 four runs. The circles show the default optimization, the squares show the
 results of optimization on odd-numbered events, the triangles show the effect
 of a different random number seed, and the inverted triangles show the effect
 of stopping the evolution at 10 generations.}
 \figlabel{top20}
\end{figure}

\section{Conclusions}\seclabel{conclusion}

We hope we have conveyed an appropriate introduction to \gp and how it can be
applied in high energy physics analyses. We have demonstrated the use of the
technique in separating the \dcs decay \kpipidcsd from the copious background
and shown that this technique can improve upon more traditional analysis
techniques. As with any analysis technique, care must be taken to
understand the possible systematic errors introduced by the technique. Finally,
we have shown that in the FOCUS case, the behavior of \mc and data is
remarkably consistent when the \gp method is applied. To our knowledge, this is
the first successful application of the \gp technique to HEP data.

Other applications for this technique are easy to imagine. Flavor tagging
(especially \bmeson mesons) has seen several successful implementations of
neural networks. \Gp may provide another means of tagging. Neutrino closure
techniques for semi-leptonic decays may also benefit from such a
technique~\cite{Frabetti:1995xq,Link:2004dh}.\footnote{Typically momentum
conservation allows for two kinematically correct solutions. Simple rules are
usually used to guess the correct solution.} 

\section*{Acknowledgments}

We wish to acknowledge the assistance of the staffs of Fermi National
Accelerator Laboratory, the INFN of Italy, and the physics departments
of the collaborating institutions. This research was supported in part
by the U.~S.  National Science Foundation, the U.~S. Department of
Energy, the Italian Istituto Nazionale di Fisica Nucleare and
Ministero dell'Istruzione dell'Universit\`a e della Ricerca, the
Brazilian Conselho Nacional de Desenvolvimento Cient\'{\i}fico e
Tecnol\'ogico, CONACyT-M\'exico, the Korean Ministry of Education, 
and the Korean Science and Engineering Foundation.

We also gratefully acknowledge our ACCRE~\cite{www_accre} colleagues, Jason H.
Moore and Bill White from the Vanderbilt Program In Human Genetics of the
Vanderbilt Medical Center, for useful discussions and assistance with the \gp
technique. 

\bibliographystyle{elsart-num}
\bibliography{physjabb,abbrev,pdg,focus,e687,theory,gp}

%\end{fmffile}
\end{document}